\documentclass[prl,twocolumn,nofootinbib,superscriptaddress]{revtex4-1}
\usepackage{graphicx}
\usepackage{dcolumn}
\usepackage{bm,bbm}
\usepackage{xcolor}
\usepackage{tcolorbox}
\usepackage{algorithm}
\usepackage{algpseudocode}
\usepackage{amsmath}
\usepackage{bbold}
\usepackage{braket}
\usepackage{nameref}
\usepackage[toc,page]{appendix}

\usepackage[colorlinks,breaklinks,linkcolor={blue},citecolor={magenta},urlcolor={blue}]{hyperref}
\usepackage{enumerate}

\makeatletter
\newcommand\org@hypertarget{}
\let\org@hypertarget\hypertarget
\renewcommand\hypertarget[2]{%
  \Hy@raisedlink{\org@hypertarget{#1}{}}#2%
  }
\makeatother

\begin{document}

\title{
Full spatial characterization of entangled structured photons}

\author{Xiaoqin Gao}
\affiliation{Nexus for Quantum Technologies, University of Ottawa, K1N 5N6, Ottawa, ON, Canada}

\author{Yingwen Zhang}
\affiliation{Nexus for Quantum Technologies, University of Ottawa, K1N 5N6, Ottawa, ON, Canada}
\affiliation{National Research Council of Canada, 100 Sussex Drive, K1A 0R6, Ottawa, ON, Canada}

\author{Alessio  D'Errico}
\affiliation{Nexus for Quantum Technologies, University of Ottawa, K1N 5N6, Ottawa, ON, Canada}

\author{Alicia  Sit}
\affiliation{Nexus for Quantum Technologies, University of Ottawa, K1N 5N6, Ottawa, ON, Canada}

\author{Khabat Heshami}
\affiliation{National Research Council of Canada, 100 Sussex Drive, K1A 0R6, Ottawa, ON, Canada}
\affiliation{Nexus for Quantum Technologies, University of Ottawa, K1N 5N6, Ottawa, ON, Canada}

\author{Ebrahim Karimi}
\affiliation{Nexus for Quantum Technologies, University of Ottawa, K1N 5N6, Ottawa, ON, Canada}
\affiliation{National Research Council of Canada, 100 Sussex Drive, K1A 0R6, Ottawa, ON, Canada}


\begin{abstract}
Vector beams (VBs) are fully polarized beams with spatially varying polarization distributions, and they have found widespread use in numerous applications such as microscopy, metrology, optical trapping, nano-photonics, and communications. The entanglement of such beams has attracted significant interest, and it has been shown to have tremendous potential in expanding existing applications and enabling new ones. However, due to the complex spatially varying polarization structure of entangled VBs (EVBs), a complete entanglement characterization of these beams remains challenging and time-consuming. Here, we have used a time-tagging event camera to demonstrate the ability to simultaneously characterize approximately $2.6\times10^6$ modes between a bi-partite EVB using only 16 measurements. This achievement is an important milestone in high-dimensional entanglement characterization of structured light, and it could significantly impact the implementation of related quantum technologies. The potential applications of this technique are extensive, and it could pave the way for advancements in quantum communication, quantum imaging, and other areas where structured entangled photons play a crucial role.
\end{abstract}

\date{\today}
\maketitle

\section*{Introduction}

Quantum entanglement~\cite{Einstein} is a remarkable feature of quantum theory that has become a cornerstone in photonic quantum information processing, encompassing quantum computation~\cite{Raussendorf2001}, quantum teleportation~\cite{Bennett1993}, quantum cryptography~\cite{Ekert}, and quantum dense coding~\cite{Bennett1992}. Demonstrated across various degrees of freedom, including polarization~\cite{PhysRevA.45.7729}, orbital angular momentum~\cite{PhysRevA.89.013829,zhang2016engineering}, path~\cite{PhysRevLett.77.1917}, and time-bin~\cite{PhysRevLett.82.2594}, photonic entanglement has gained prominence as technologies leveraging it become more integrated into daily life. Consequently, examining intricate entanglement structures and their applications is increasingly vital.

Distinct from homogeneously polarized light, vector beams (VBs) display a complex, spatially varying polarization structure~\cite{zhan2009cylindrical}. Owing to their unique attributes, VBs have recently attracted attention in diverse fields such as nonlinear optics~\cite{PhysRevLett.90.013903}, high-resolution imaging~\cite{Chen:07}, laser processing~\cite{Kozawa:10}, classical or quantum communication \cite{d2012complete,farias2015resilience, ndagano2018creation}, sharper focus spots~\cite{PhysRevLett.91.233901,wang2008creation}, and quantum key distribution~\cite{souza2008quantum,PhysRevLett.113.060503}. Two primary approaches to generate VBs include the intra-cavity method, which involves inserting mode-selection optical elements in a laser resonator~\cite{sakai2007optical, miyai2006lasers,Iwahashi:11}, and the extra-cavity method, which entails transforming Gaussian beams into VBs using specialized optical elements like $q$-plates~\cite{Cardano:12, Cardano:13} or a spatial light modulator combined with a Sagnac interferometer~\cite{fickler2013real}.

Entangled VBs (EVBs) have been demonstrated in~\cite{PhysRevA.94.030304, PhysRevResearch.2.043350}, where polarization-entangled photon pairs are converted into VBs, transforming spatially homogeneous polarization entanglement into a complex, spatially variant polarization entanglement structure. However, thoroughly characterizing this spatially dependent entanglement is challenging. Employing single-pixel detectors, such as avalanche photodiodes, necessitates $(N\times N)^2 \times 16$ measurements, with $N\times N$ representing the number of spatial locations on each beam to be probed, and 16 denoting the required polarization measurement combinations~\cite{Altepeter2005}. By using a triggerable intensified CCD (ICCD) camera, where a single-pixel detector triggers the ICCD, the number of measurements can be reduced to $(N\times N) \times 16$, as partially demonstrated in the case of a homogeneously polarized beam entangled to a VB~\cite{fickler2013real, PhysRevA.89.060301}.

Here, we present the full characterization of spatially dependent polarization entanglement between EVBs using only 16 polarization measurements, facilitated by a time-tagging event camera. Covering approximately $40\times40$ pixels per beam, $\sim2.6\times10^6$ modes between a bi-partite entangled state are measured simultaneously. To showcase the versatility of our technique, we perform the characterization for various polarization topological patterns, including vector-vortex, lemon, star, and spiral polarization patterns, all exhibiting exceptional agreement with the anticipated entanglement pattern derived from theory.

\section*{Results}
\begin{figure*}[t]
\includegraphics [width= 0.8\textwidth]{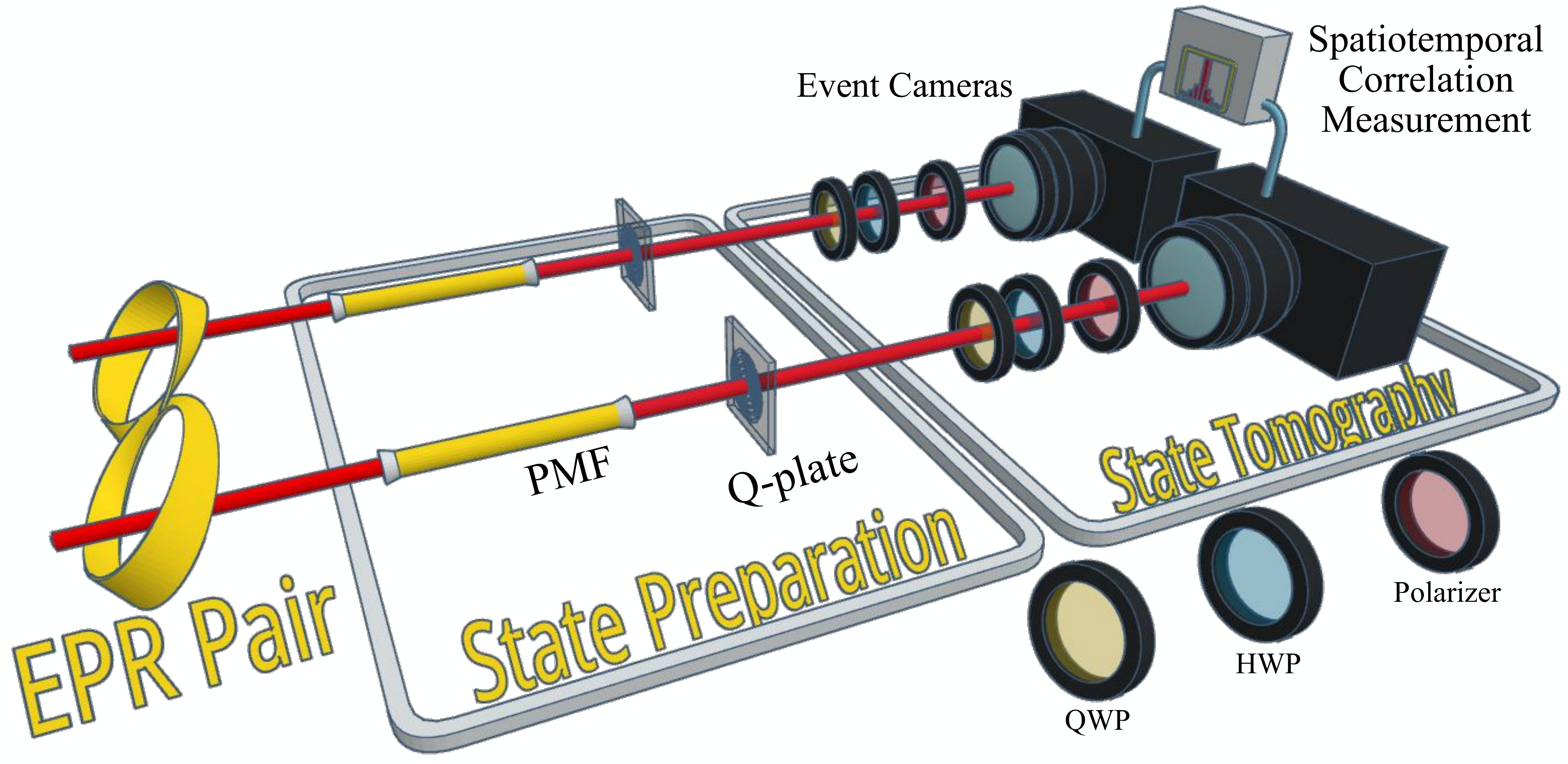}
\centering
\caption{
{\bf Conceptual setup for the generation and characterization of vector beam entanglement.} Einstein–Podolsky–Rosen (EPR) photon pairs entangled in the polarization degree of freedom are first spatially filtered by polarization-maintaining single-mode fibres (PMF) into the fundamental mode before being sent onto the $q$-plates (with different topological charges) to create entangled structured photons. Polarization state tomography is performed using a combination of quarter-wave plates (QWP), half-wave plates (HWP), and polarizers with the photon pairs detected by time-tagging event cameras from which spatial-temporal information of the photons is collected. A spatial-temporal correlation measurement is then performed for each of the 16 combinations of polarization measurement. The density matrix and, subsequently, the Bell states for each pixel combination can then be determined.}
\label{fig:concept}
\end{figure*}
\noindent\textbf{Theory:} The conceptual diagram for the generation and characterization of spatially varying polarisation entangled photons is shown in Fig.~\ref{fig:concept}. A polarization-entangled state with a Gaussian beam profile of the following form is first prepared,
\begin{align}\label{Bell}
    \ket{\Psi_0}=\frac{1}{\sqrt{2}}\left(\ket{H,V}-\ket{V,H}\right)=\frac{i}{\sqrt{2}}\left(\ket{L,R}-\ket{R,L}\right),
\end{align}
where $\ket{L,R}=\ket{L}_s\otimes\ket{R}_i$ with $\ket{L}$ and $\ket{R}$ denoting left- and right-circular polarization, and the subscripts $s$ and $i$ denoting the signal and idler photons, respectively. 

The unitary action of a $q$-plate with topological charge $q$ acting on a circularly polarized Gaussian beam can be written as~\cite{karimi2009light},
\begin{eqnarray} \label{qplate}
    \hat{U}_q^\delta\cdot\left[ \begin{array}{ll}
        \ket{L}\\
        \ket{R}\\
    \end{array} \right]&=& F_{0}(r)\cos\left(\frac{\delta}{2}\right)\left[ \begin{array}{ll}
        \ket{L}\\
        \ket{R}\\
    \end{array} \right]\\ \nonumber
    &+&i\sin\left(\frac{\delta}{2}\right)F_{q}(r)\left[ \begin{array}{ll}
       e^{-i2q\theta}\ket{R}\\
      e^{i2q\theta}\ket{L}\\
    \end{array} \right],
\end{eqnarray}
where $F_0(r)$ and $F_q(r)$ determine the radial profiles of the unconverted and converted beams, respectively -- here, we considered $q$-plates wherein the local optical axis distribution of the liquid crystals is $\alpha(\theta)=q\theta$. Both $F_0(r)$ and $F_q(r)$ can be expressed analytically in terms of hypergeometric Gauss modes \cite{karimi2007hypergeometric}. However, for $q$-plates with low $q$ values, within a good approximation, both radial functions can be expressed in terms of the Laguerre-Gauss modes possessing azimuthal indices of $0$ and $2q$, and radial index zero at $z=0$, respectively, i.e., $F_{q}(r)e^{\pm i2q\theta}\equiv\text{LG}_{0,\pm 2q}(r,\theta)$~\cite{karimi2009light}. $\delta \in [0,\pi]$ is the optical retardation of the $q$-plate, which can be tuned based on the voltage applied to the $q$-plates~\cite{slussarenko2011tunable}. Therefore, sending each photon onto different $q$-plates yields the following complex space-varying polarization-entangled state,
\begin{equation}
    \ket{\Psi_{f}}=\frac{1}{\sqrt{2}}\left(\hat{U}_{q_s}^{\delta_s}\cdot\ket{L}_s\hat{U}_{q_i}^{\delta_i}\cdot\ket{R}_i-\hat{U}_{q_s}^{\delta_s}\cdot\ket{R}_s\hat{U}_{q_i}^{\delta_i}\cdot\ket{L}_i\right),
\end{equation}
here, $\hat{U}_{q_s}^{\delta_s}$ and $\hat{U}_{q_i}^{\delta_i}$ denotes the signal and idler $q$-plate actions, respectively -- note that the global U(1)-phase of $\pi/2$ is omitted. The four parameters of $\{q_s,\delta_s,q_i,\delta_i\}$ determine the polarization entanglement state between signal and idler photons. For instance, setting $q_s=q_i=1/2$ and $\delta_s=\delta_i=\pi/2$, the polarization Einstein–Podolsky–Rosen (EPR) state is transformed into the radial and azimuthal entangled state, i.e., $\left(\ket{\text{radial},\text{azimuthal}}-\ket{\text{azimuthal},\text{radial}}\right)/\sqrt{2}$. To characterize the spatially varying entanglement between the entangled photons, polarization state tomography is performed using 16 different combinations of polarization measurement. Each photon is then collected by a time-tagging event camera from which spatial-temporal information about the photons is recorded. A spatial-temporal correlation measurement is then performed between the two cameras for each of the 16 polarization measurements from which the spatially varying density matrix and, subsequently, the Bell states contributions can be determined.\\

\begin{figure*}
\includegraphics [width= 0.96\textwidth]{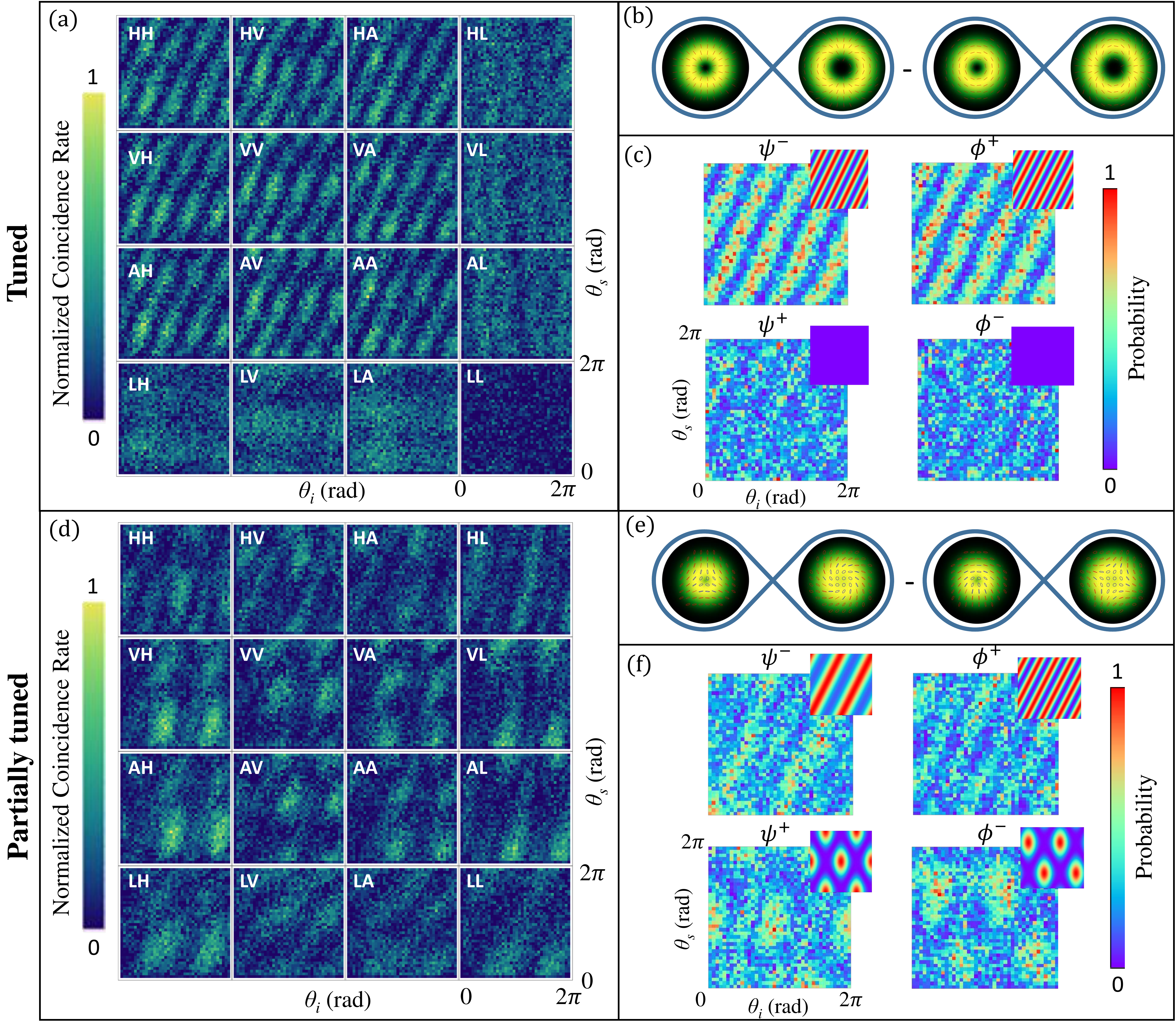}
\centering
\caption{\textbf{Measured spatial correlations and entanglement characterization for bi-photon with spatially varying polarization states generated via two $q$-plates.} (a) Two-photon spatial correlations between two entangled vector beams measured in the azimuthal degree of freedom from 16 different polarization measurements. Two tuned $q$-plates ($\delta_s=\delta_i=\pi$) with topological charges $q_i = 1$ and $q_s = 1/2$ are used to generate the entangled vector vortex photons.
(b) Graphical representation of the polarization patterns overlaid on the photon probability distribution for an input EPR pair in the antisymmetric Bell state.
(c) Reconstructed probabilities for the four Bell states $\{\ket{\psi^+},\ket{\psi^-},\ket{\phi^+},\ket{\phi^-}\}$. The insets are the respective theoretical probabilities. (d-f) Corresponding results for two partially tuned $q$-plates ($\delta_s=\delta_i=\pi/2$) with topological charges $q_i = 1$ and $q_s = 1/2$. Red lines in  (b) and red (blue) ellipses in (e) show the major axes of the polarization.}
\label{prob1&1/2}
\end{figure*}

\begin{figure*}
\includegraphics [width=1\linewidth]{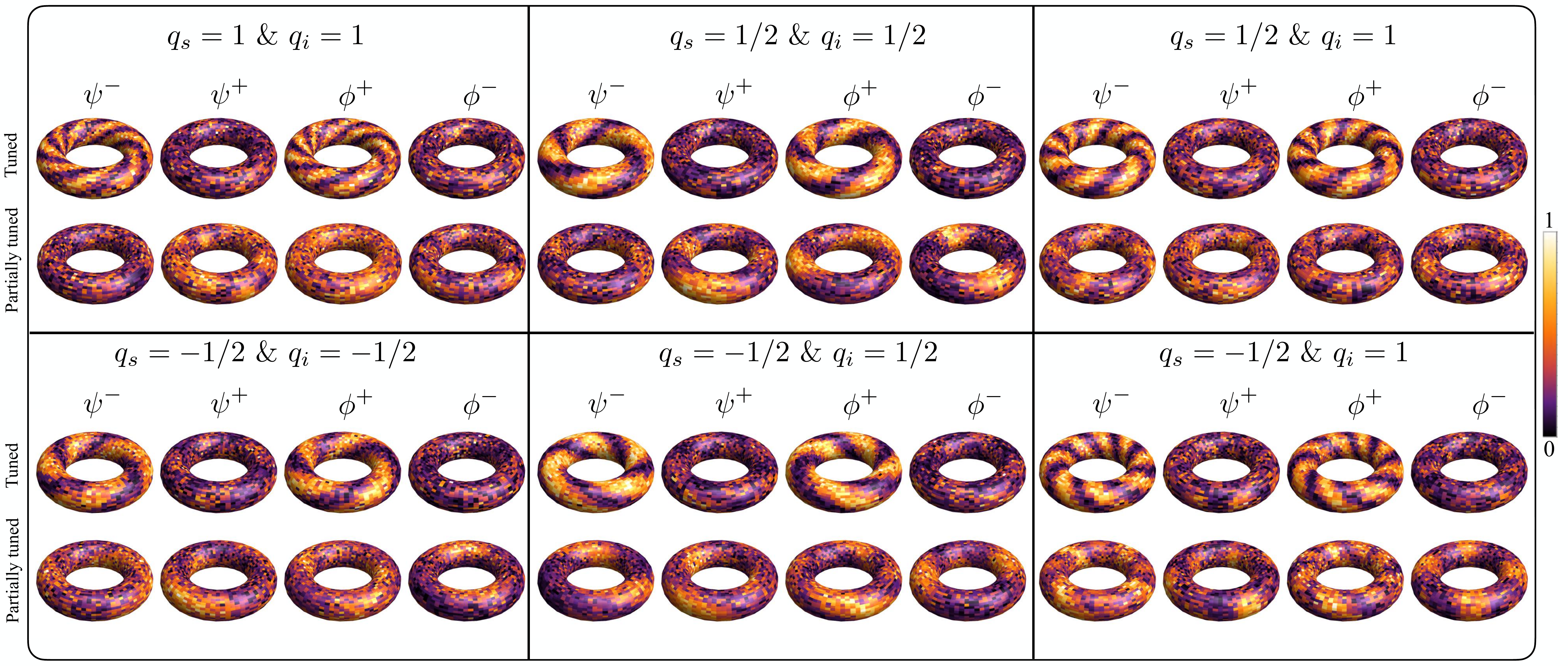}
\centering
\caption{\textbf{The experimental results for the probabilities of the four Bell states.} The Bell state probabilities at different azimuthal angles between the two entangled vector beams are mapped onto a torus. The vector beams are generated for a combination of $q$-plates with a fractional topological charge of $q=-1/2$, $1/2$, or $1$ in each path.}
\label{3D}
\end{figure*}

\noindent\textbf{Experimental Results:} In this work, we demonstrate the entanglement characterization for two $q$-plate settings: the case of when both $q$-plates are perfectly tuned ($\delta_s=\delta_i=\pi$) and when both are partially tuned ($\delta_s=\delta_i=\pi/2$). For two tuned $q$-plates with topological charges $q_s$ and $q_i$ applied to the signal and idler photons, respectively, the transverse position-dependent polarization entanglement between the EVBs can be decomposed into the four Bell states, $\ket{\phi^{\pm}} =\frac{1}{\sqrt{2}}(\ket{H}_s\ket{H}_i \pm \ket{V}_s\ket{V}_i)$ and $\ket{\psi^{\pm}} =\frac{1}{\sqrt{2}}(\ket{H}_s\ket{V}_i \pm \ket{V}_s\ket{H}_i)$, with the following contributions,
\begin{equation} \label{protuned}
\begin{split}
P_{\phi^{+}}^\text{t}&=F^2_{q_s,q_i}(r_s,r_i)\sin^2 2(q_s \theta_s-q_i \theta_i),\\
P_{\phi^{-}}^\text{t}&=0,\\
P_{\psi^{+}}^\text{t}&=0,\\
P_{\psi^{-}}^\text{t}&=F^2_{q_s,q_i}(r_s,r_i)\cos^2 2(q_s \theta_s-q_i \theta_i),
\end{split}
\end{equation}
where $F_{q_s,q_i}(r_s,r_i)=F_{q_s}(r_s)F_{q_i}(r_i)$, and $(r_s,\theta_s)$ and $(r_i,\theta_i)$ are the transverse coordinates of signal and idler photons. For the partially tuned case, ($\delta_s=\delta_i=\pi/2$), the contributions are,
\begin{align}
    P_{\phi^+}^\text{pt} &= \frac{1}{4}F^2_{q_s,q_i}(r_s,r_i)\sin^2 2(q_s \theta_s-q_i \theta_i), \nonumber\\
    P_{\phi^-}^\text{pt} &= \frac{1}{4}[F_{0,q_i}(r_s,r_i)\sin{2q_i\theta_i} - F_{q_s,0}(r_s,r_i) \sin{2q_s\theta_s}]^2, \nonumber\\
    P_{\psi^+}^\text{pt} &= \frac{1}{4}[F_{0,q_i}(r_s,r_i)\cos{2q_i\theta_i} - F_{q_s,0}(r_s,r_i) \cos{2q_s\theta_s}]^2, \nonumber\\
    P_{\psi^-}^\text{pt} &= \frac{1}{4}\left[F_{0,0}(r_s,r_i)-F_{q_s,q_i}(r_s,r_i)\sin{2(q_s \theta_s-q_i \theta_i})\right]^2.
    \label{prodetuned}
\end{align}
Detailed calculations can be found in the Supplementary materials.

The Bell state contributions for EVBs generated with tuned $q$-plates depend only on the azimuthal degree of freedom (DOF), $\theta_s$ and $\theta_i$, and of course, the plates, topological charges $q_s, q_i$, while both $\phi^{-}$ and $\psi^{+}$ states make no contributions. For partially tuned $q$-plates, the situation is more complex since all four Bell states make contributions. Indeed, Bell states are a function of both the azimuthal DOF, $\theta_s$ and $\theta_i$, and the radial DOF, $r_s$ and $r_i$.

In the experimental demonstration, a single time-tagging camera was employed as an alternative to the originally proposed dual-camera setup. Each beam of the EVB pair was directed towards a corner of the camera, encompassing an area of roughly $40\times 40$ pixels. An initial temporal correlation measurement was conducted to identify photon pairs using the position and time data of each photon captured by the camera. This was followed by a spatial correlation measurement conducted in cylindrical polar coordinates. This procedure was repeated for 16 distinct polarization measurement combinations, as required for bi-photon polarization state tomography~\cite{Altepeter2005}. The quantum state tomography for EVBs generated with $q$-plates of topological charges $q_s = 1/2$ and $q_i= 1$ is illustrated in Fig.~\ref{prob1&1/2}. The measured spatial correlation between signal and idler photons in the azimuthal DOF $\theta$, with radial DOF $r$ summed over, is presented for the 16 different polarization combinations employed in polarization state tomography, see Fig.~\ref{prob1&1/2}(a). Correlations in the radial DOF exhibited minimal variation between distinct polarization measurements and are, therefore, not depicted here, but can be found in the Supplementary materials. Subsequently, an azimuthal angle-dependent density matrix was reconstructed, and to simplify visualization, the corresponding Bell state decomposition is displayed in Fig.~\ref{prob1&1/2}(c). We observe azimuthally that the measured Bell state contributions oscillate between the $\ket{\psi^-}$ and $\ket{\phi^+}$ states, with negligible contributions from the $\ket{\phi^-}$ and $\ket{\psi^+}$ states. This outcome aligns well with the theoretical calculations reported in Eq.~\eqref{protuned}. Imperfections in the $q$-plates, entanglement, and some noise may contribute to residual oscillations in the $\ket{\psi^+}$ and $\ket{\phi^-}$ states. The measured concurrence, averaged over all $\ket{\phi}$, of the EVB is $0.540\pm 0.005$. Results for two partially tuned $q$-plates with topological charges $q_s= 1/2$ and $q_i= 1$ are displayed in Fig.~\ref{prob1&1/2}(d-f). Due to the increased complexity of the spatially variant structure and sensitivity to optical alignment, the state tomography results are noisier, yet they still demonstrate reasonable agreement with theoretical expectations. The average concurrence of the EVB is measured to be $0.517\pm 0.005$.

Figure~\ref{3D} displays the experimentally measured Bell state decomposition for EVBs generated using both tuned and partially tuned $q$-plates at various combinations of topological charges. The probability amplitudes of the Bell states are mapped onto a torus due to the cyclic nature of the azimuthal angle of polar coordinates $(\theta_s,\theta_i)$. For the measured spatial correlations of all $q$-plate combinations, please refer to the Supplementary materials.

One significant limitation of the current demonstration is the low efficiency of single-photon detection, i.e. the camera, which results in significant noise and extended data acquisition times of eight minutes per polarization measurement. The camera's low efficiency of approximately 8\%~\cite{Vidyapin2022}, combined with the need to couple polarization-entangled photons into single-mode fibers with a coupling efficiency of around 30\%, only provides a coincidence efficiency of 0.06\%. To improve the system's performance, a potential first step would be to generate entangled photons directly from fibers through spontaneous four-wave mixing~\cite{Takesue2004}, eliminating the losses associated with fiber coupling. This modification could result in an order of magnitude improvement in coincidence efficiency and, consequently, a reduction in data acquisition time. Moreover, with the coincidence efficiency scaling quadratically relative to the single-photon efficiency, further improvements could be achievable with the development of more sensitive cameras in the near future, allowing for entanglement characterization to be performed in mere seconds.

\section*{Conclusion and outlook}

In this study, we have successfully demonstrated the generation and characterization of non-uniform transverse polarization entanglement between various classes of vector beams in a scan-free manner. The demonstrated method can be invaluable for quickly characterizing spatially varying entanglement in EVBs, which are gaining interest in various high-dimensional quantum systems such as quantum communication and quantum imaging. We utilized spatial-temporal correlation measurements on data collected by a time-tagging event camera to perform polarization state tomography on approximately $2.6\times10^6$ ($40\times40$) spatial modes using only 16 polarization measurements. The measurement results agreed well with theoretical predictions. Future improvements to the detection efficiency of time-tagging camera technology could potentially enable real-time state tomography on EVBs by splitting the beams onto multiple cameras, with each camera measuring a different polarization. These developments would significantly enhance the capabilities of quantum communication and quantum imaging systems. The local entanglement measurements in our system could be exploited to enhance the sensitivity of photonic gears \cite{d2013photonic, barboza2022ultra}. Our study also shows some promising applications, for example, quantum sensing of birefringent materials, and the generation of high-dimensional topological structures such as 4D-skyrmions.

\bibliographystyle{apsrev4-1fixed_with_article_titles_full_names_new}
\bibliography{refs}


\vspace{0cm}
\vspace{1 EM}

\noindent\textbf{Acknowledgments}
\noindent The authors would like to thank Ryan Hogan for helping with coding during the data analysis, and Paul Corkum, Felix Hufnagel, Dilip Paneru, and Fr\'ed\'eric Bouchard for their valuable discussions. K.H. acknowledges that the National Research Council of Canada (NRC) headquarters is located on the traditional unceded territory of the Algonquin Anishinaabe and Mohawk people. This work was supported by the Canada Research Chairs (CRC), the High Throughput and Secure Networks (HTSN) Challenge Program at the National Research Council of Canada, and the Joint Centre for Extreme Photonics (JCEP). 
\vspace{1 EM}

\noindent\textbf{Author Contributions}
E.K. conceived the idea. X.G. and Y.Z. devised the experimental setup. X.G. performed the experiment. A.S. and A.D. fabricated $q$-plates. X.G., Y.Z., and A.D. analyzed the data. K.H., P.C. and E.K. supervised the project. All authors contributed to the writing of the manuscript.
\vspace{1 EM}

\noindent\textbf{Data availability}
\noindent
The data that support the findings of this study are available from the corresponding author upon reasonable request.
\vspace{1 EM}

\noindent\textbf{Code availability}
\noindent
The code used for the data analysis is available from the corresponding author upon reasonable request.

\vspace{1 EM}
\noindent\textbf{Ethics declarations} The authors declare no competing interests.

\vspace{1 EM}
\noindent\textbf{Corresponding authors}
Correspondence and requests for materials should be addressed to yzhang6@uottawa.ca.

\clearpage

\section*{Methods}
\begin{figure}[t]
\includegraphics [width= 1\linewidth]{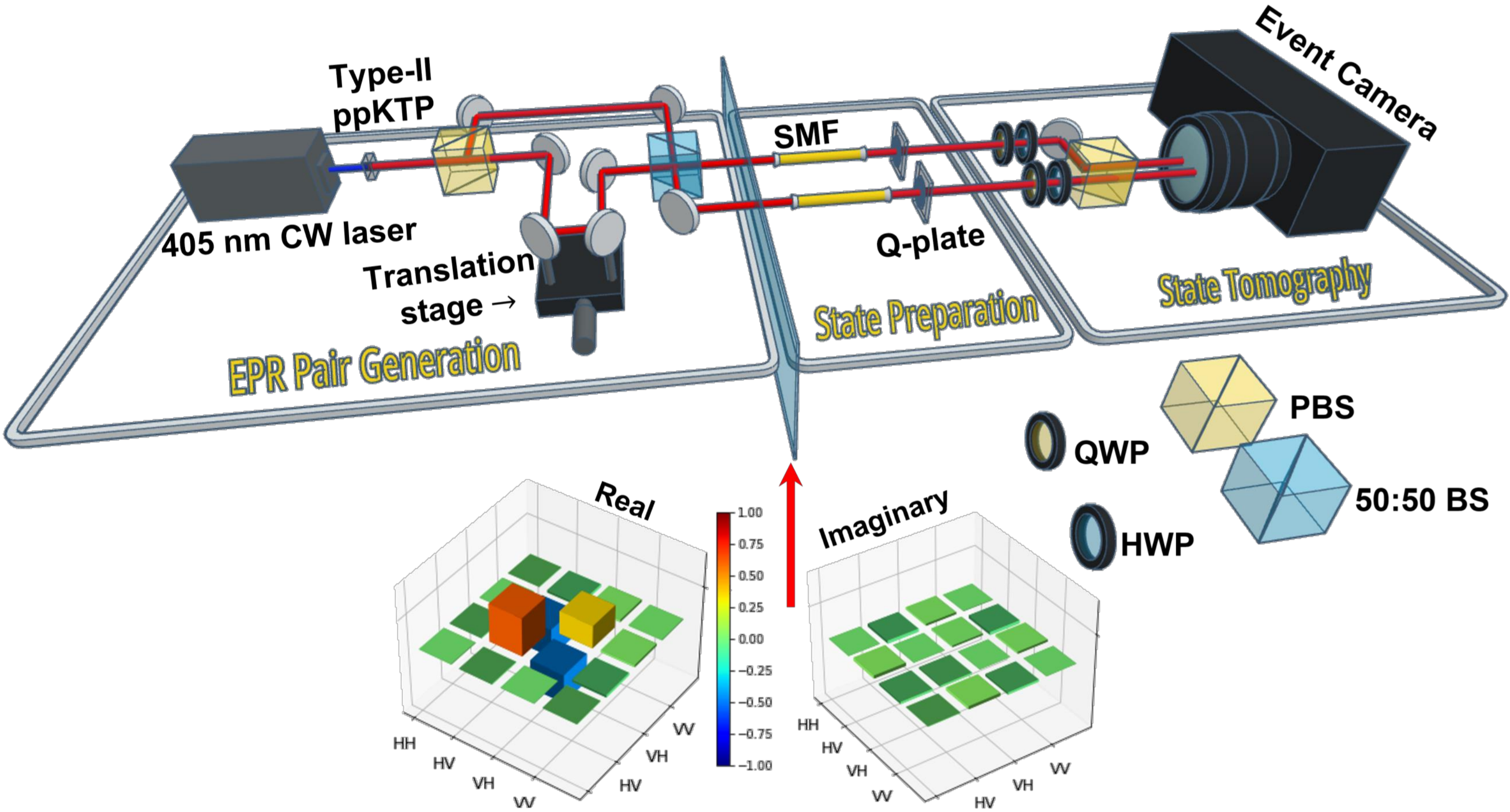}
\centering
\caption{{\bf Experimental setup for generation and characterization of entangled vector beams.} EPR Pair Generation - A Type-II ppKTP crystal is pumped by a 405\,nm continuous-wave laser to generate photon pairs orthogonal in polarization through SPDC. An antisymmetric Bell state $\ket{\psi^-}$ is generated through correlation measurements between the two exit ports of a Hong-Ou-Mandel interferometer. State Preparation - The polarization-entangled photon pair are first coupled into single-mode fibers (SMFs) for spatial mode filtering. $q$-plates are then used to transform the spatially uniform polarization entanglement into entangled vector beams. State Tomography - Polarization state tomography is performed using a series of a quarter-wave plate (QWP), a half-wave plate (HWP), and a Polarizing beamsplitter (PBS) after which each photon of the entangled pair is guided to hit a different region of an event camera where spatial and temporal information of the photons are recorded. The inset below the experimental setup shows the polarization density matrix (real and imaginary) measured after the EPR Pair Generation stage with the entanglement being in the $\ket{\psi^-}$ Bell state.}
\label{fig:setup}
\end{figure}

The experimental setup for generation and characterization of entangled vector beams is shown in Fig.~\ref{fig:setup}. The setup can be divided into three stages: EPR Pair Generation, State Preparation, and State Tomography. In the EPR Pair Generation stage, a $5$-mm-thick Type-II periodically poled potassium titanyl phosphate (ppKTP) crystal is pumped by a 405\,nm continuous-wave laser to generate photon pairs with orthogonal polarization through the process of spontaneous parametric down-conversion (SPDC). The photon pairs then go through two-photon interference in a Hong-Ou-Mandel interferometer which will act as a filter for the antisymmetric $\ket{\psi^-}$ Bell state,
\begin{equation}
    \ket{\Psi_0} =\frac{1}{\sqrt{2}}(\ket{H}_s\ket{V}_i-\ket{V}_s\ket{H}_i),
\end{equation}

\noindent when time correlation measurements between the two exit ports of the interferometer are made. Performing polarization state tomography at this stage between the two exit ports precisely gives the $\ket{\psi^-}$ Bell state where the reconstructed density matrix can be seen in the inset of Fig.~\ref{fig:setup} with a concurrence and purity of $\sim 99\%$. In the State Preparation stage, the polarization entangled photon pair is first coupled into single-mode fibers (SMFs) for spatial mode filtering leaving only the Gaussian mode on exit while still maintaining polarization entanglement. $q$-plates are then used to transform the spatially uniform polarization entanglement into entangled vector beams with spatially varying polarization entanglement. Lastly, in the State Tomography stage, polarization state tomography is performed through 16 different polarization measurements using a series of a quarter-wave plate, half-wave plate, and polarizing beamsplitter after which each photon of the entangled pair is guided to hit a different region of a time-tagging event camera (TPX3CAM~\cite{ASI, Nomerotski2019}) where spatial and temporal information of the photons are recorded from which spatial-temporal correlation measurements are made.

\clearpage
\onecolumngrid
\renewcommand{\figurename}{\textbf{Figure}}
\setcounter{figure}{0} \renewcommand{\thefigure}{\textbf{S{\arabic{figure}}}}
\setcounter{table}{0} \renewcommand{\thetable}{S\arabic{table}}
\setcounter{section}{0} \renewcommand{\thesection}{Section S\arabic{section}}
\setcounter{equation}{0} \renewcommand{\theequation}{S\arabic{equation}}
\onecolumngrid

\begin{center}
{\Large Supplementary Materials for: \\ Full spatial characterization of entangled structured photons}
\end{center}
\vspace{1 EM}

\section{Two tuned $q$-plates}
The unitary action $\hat{U}_q^\delta$ of a $q$-plate with topological charge $q$ acting on a circularly polarized Gaussian beam to first-order approximation can be written as
\begin{equation} \label{eqa1}
    \hat{U}_q^\delta\cdot\left[ \begin{array}{ll}
        \ket{L}\\
        \ket{R}\\
    \end{array} \right] = F_{0}(r)\cos\left(\delta/2\right)\left[ \begin{array}{ll}
        \ket{L}\\
        \ket{R}\\
    \end{array} \right]
   + i \sin\left(\delta/2\right)F_{q}(r)\left[ \begin{array}{ll}
       e^{-i2q\theta}\ket{R}\\
      e^{i2q\theta}\ket{L}\\
    \end{array} \right],
\end{equation}
where $\ket{L}$ and $\ket{R}$ are left- and right-circular polarizations respectively. 
$F_{q}(r)e^{\pm i2q\theta}\equiv\text{LG}_{0,\pm 2q}(r,\theta)$ is the Laguerre-Gauss mode with orbital angular momentum index $2q$ and radial index zero at $z=0$. $\delta \in [0,\pi]$ is an adjustable parameter that can be tuned based on the voltage applied to the $q$-plate. The $q$-plates are considered tuned with $\delta=\pi$, and when $\delta=\pi/2$, the $q$-plates are considered partially tuned.

In our work, we first prepare an anti-symmetric entangled state in the polarization degree of freedom (DOF),
\begin{equation} \label{eqa2}
\begin{split}
\ket{\Psi_{0}} & =\frac{1}{\sqrt{2}}(\ket{H}_s\ket{V}_i-\ket{V}_s\ket{H}_i)\\
    & = \frac{i}{\sqrt{2}}(\ket{L}_s\ket{R}_i-\ket{R}_s\ket{L}_i),
\end{split}
\end{equation}
with $s$ and $i$ denoting the signal and idler photons respectively. 

Applying two tuned $q$-plates ($\delta_s=\delta_i=\pi$) with topological charges $q_s$ and $q_i$ respectively to each photon gives the following final state,
\begin{equation} \label{eqa3}
\ket{\Psi_{f}} =\frac{i}{\sqrt{2}}F_{q_s,q_i}(r_s,r_i)\left[ e^{-2i (q_s \theta_s-q_i \theta_i)}\ket{R}_s \ket{L}_i-  e^{2i (q_s \theta_s-q_i \theta_i)}\ket{L}_s \ket{R}_i\right],
\end{equation}
where $F_{q_s,q_i}(r_s,r_i)=F_{q_s}(r_s)F_{q_i}(r_i)$. 

Rewriting in the horizontal ($H$) and vertical ($V$) polarization basis gives
\begin{equation} \label{eqa3}
\ket{\Psi_{f}} =F_{q_s,q_i}(r_s,r_i)\left[\sin 2(q_s \theta_s-q_i \theta_i)\ket{\phi^+} - \cos 2(q_s \theta_s-q_i \theta_i)\ket{\psi^-}\right].
\end{equation}
where $\ket{\phi^+} = \frac{1}{\sqrt{2}}(\ket{H}_s \ket{H}_i + \ket{V}_s \ket{V}_i$) and $\ket{\psi^-} = \frac{1}{\sqrt{2}}(\ket{H}_s \ket{V}_i - \ket{V}_s \ket{H}_i)$ are two of the four Bell states.

Therefore, the probabilities for the final state to be in one of the four Bell states are as follows,
\begin{align}
    P_{\phi^+}^\text{t} &= F^2_{q_s,q_i}(r_s,r_i)\sin^2 2(q_s \theta_s-q_i \theta_i) \nonumber\\
    P_{\phi^-}^\text{t} &= 0 \nonumber\\
    P_{\psi^+}^\text{t} &= 0 \nonumber\\
    P_{\psi^-}^\text{t} &= F^2_{q_s,q_i}(r_s,r_i)\cos^2 2(q_s \theta_s-q_i \theta_i)
\end{align}

\section{Two partially tuned $q$-plates}

When using two partially tuned $q$-plates ($\delta=\pi/2$), the initial entangled state from Eq.\eqref{eqa2} becomes,
\begin{align} \label{eqb2}
\ket{\Psi_{f}} & = \frac{i}{2\sqrt{2}} \Bigl\{(F_{0,0}(r_s,r_i)-F_{q_s,q_i}(r_s,r_i)e^{2i(q_s \theta_s-q_i \theta_i})\ket{L}_s\ket{R}_i\nonumber\\
&\quad - \bigl[F_{0,0}(r_s,r_i)-F_{q_s,q_i}(r_s,r_i)e^{-2i(q_s \theta_s-q_i \theta_i}\bigr]\ket{R}_s\ket{L}_i\nonumber\\
&\quad + \bigl[F_{0,q_i}(r_s,r_i)(r_i)e^{2iq_i\theta_i} - F_{q_s,0}(r_s,r_i) e^{2iq_s\theta_s}\bigr]\ket{L}_s\ket{L}_i\nonumber\\
&\quad - \bigl[F_{0,q_i}(r_s,r_i)(r_i)e^{-2iq_i\theta_i} - F_{q_s,0}(r_s,r_i) e^{-2iq_s\theta_s}\bigr]\ket{R}_s\ket{R}_i\Bigr\}.
\end{align}

Rewriting in the $H$ and $V$ basis gives
\begin{align} \label{eqb2}
\ket{\Psi_{f}} & = \frac{1}{2}\Bigl\{-F_{q_s,q_i}(r_s,r_i)\sin{2(q_s \theta_s-q_i \theta_i})\ket{\phi^+}\nonumber\\
&\quad + \bigl[F_{0,0}(r_s,r_i)-F_{q_s,q_i}(r_s,r_i)\sin{2(q_s \theta_s-q_i \theta_i}\bigr]\ket{\psi^-}\nonumber\\
&\quad + \bigl[F_{0,q_i}(r_s,r_i)\sin{2q_i\theta_i} - F_{q_s,0}(r_s,r_i) \sin{2q_s\theta_s}\bigr]\ket{\phi^-}\nonumber\\
&\quad + \bigl[F_{0,q_i}(r_s,r_i)\cos{2q_i\theta_i} - F_{q_s,0}(r_s,r_i) \cos{2q_s\theta_s}\bigr]\ket{\psi^+}\Bigr\},
\end{align}
with $\ket{\phi^\pm} = \frac{1}{\sqrt{2}}(\ket{H}_s \ket{H}_i \pm \ket{V}_s \ket{V}_i$), and $\ket{\psi^\pm} = \frac{1}{\sqrt{2}}(\ket{H}_s \ket{V}_i \pm \ket{V}_s \ket{H}_i)$ being the four Bell states.

Thus, the probabilities for the final state to be in one of the four Bell states are as follows,
\begin{align}
    P_{\phi^+}^\text{pt} &= \frac{1}{4}F^2_{q_s,q_i}(r_s,r_i)\sin^2 2(q_s \theta_s-q_i \theta_i) \nonumber\\
    P_{\phi^-}^\text{pt} &= \frac{1}{4}[F_{0,q_i}(r_s,r_i)\sin{2q_i\theta_i} - F_{q_s,0}(r_s,r_i) \sin{2q_s\theta_s}]^2 \nonumber\\
    P_{\psi^+}^\text{pt} &= \frac{1}{4}[F_{0,q_i}(r_s,r_i)\cos{2q_i\theta_i} - F_{q_s,0}(r_s,r_i) \cos{2q_s\theta_s}]^2 \nonumber\\
    P_{\psi^-}^\text{pt} &= \frac{1}{4}\left[F_{0,0}(r_s,r_i)-F_{q_s,q_i}(r_s,r_i)\sin{2(q_s \theta_s-q_i \theta_i})\right]^2.
\end{align}

\section{Experimental results}

\begin{figure*}[tbph]
\includegraphics [width= 0.9\textwidth]{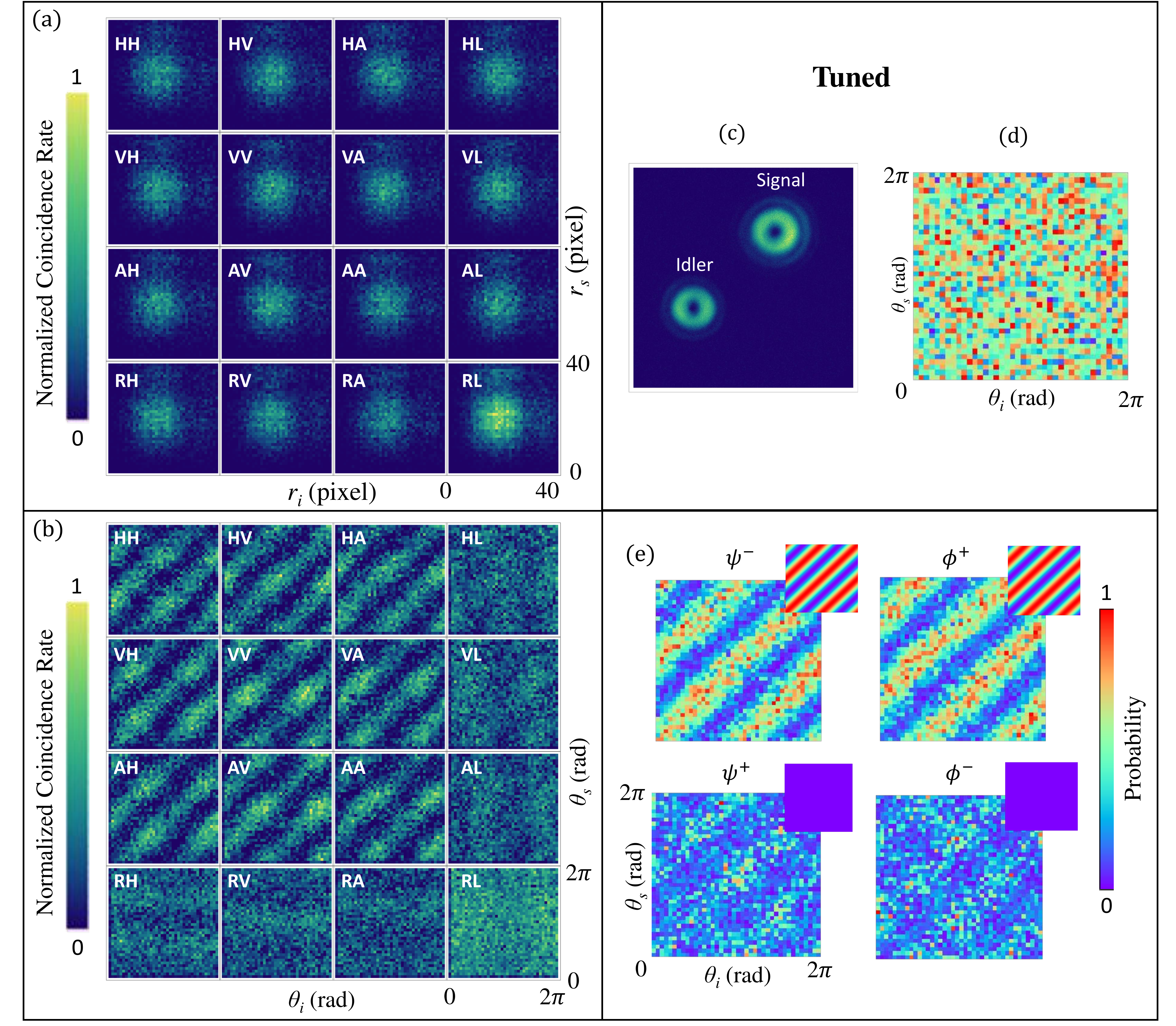}
\centering
\caption{\textbf{Measured spatial correlations and entanglement characterization for bi-photon with spatially varying polarization states generated via two $q$-plates.} 
(a)-(b) Two-photon spatial correlations between two entangled vector beams
measured, in the radial and azimuthal DOF respectively, from 16 different polarization measurements.  
Two tuned $q$-plates ($\delta_s=\delta_i=\pi$) with topological charges $q_i = 1/2$ and $q_s = 1/2$ are used to generate the entangled vector vortex photons.
(c) Continuous exposure images recorded on the camera.
(d) The concurrence of the entangled state and the average concurrence is measured to be $0.566\pm0.035$.
(e) Reconstructed probabilities for the four Bell states $\{\ket{\psi^+},\ket{\psi^-},\ket{\phi^+},\ket{\phi^-}\}$. The insets are the respective theoretical probabilities.
}
\label{correlation1/2&1/2}
\end{figure*}

\begin{figure*}[tbph!]
\includegraphics [width=0.9\linewidth]{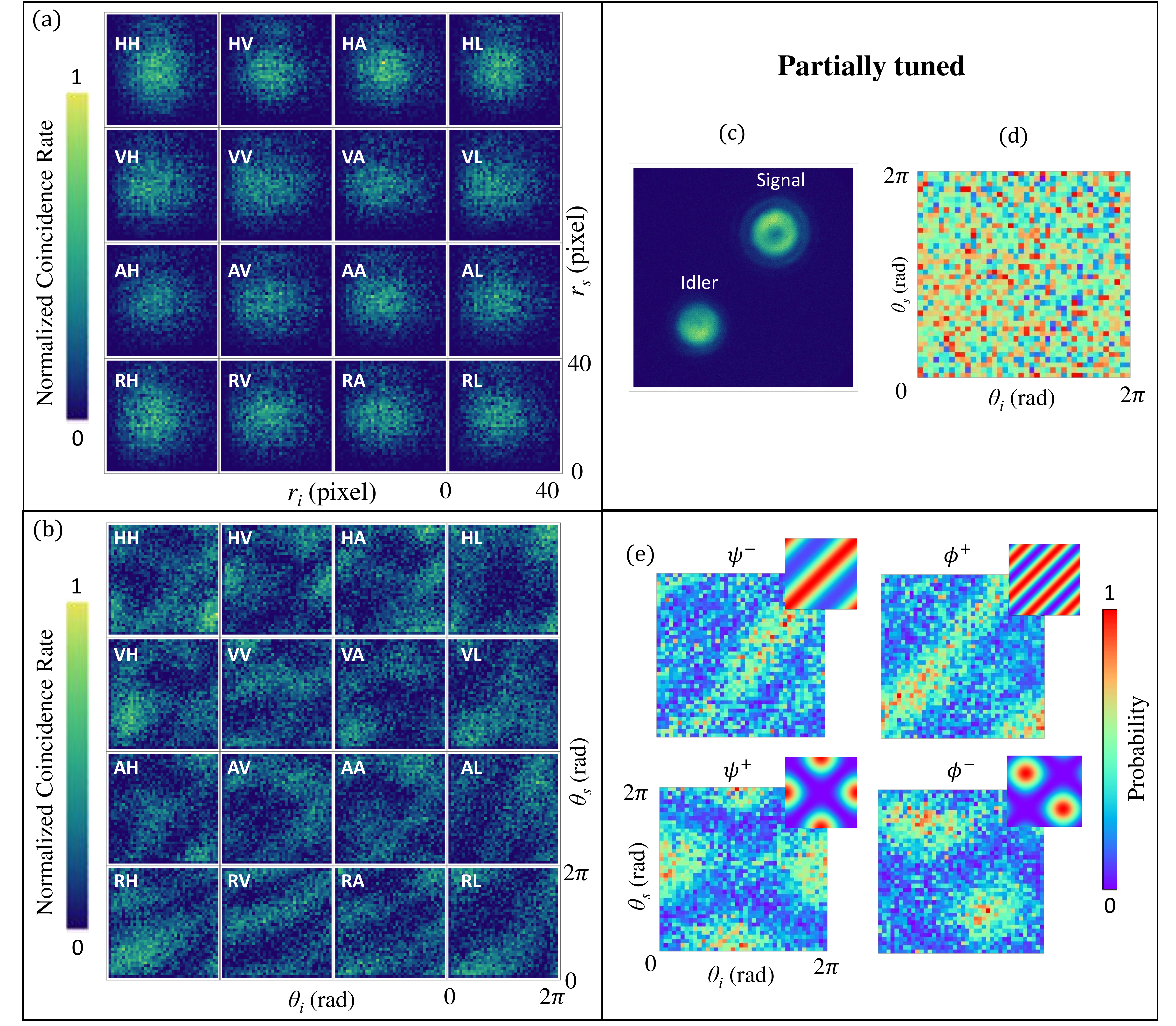}
\centering
\caption{\textbf{Measured spatial correlations and entanglement characterization for bi-photon with spatially varying polarization states generated via two $q$-plates.} 
(a)-(b) Two-photon spatial correlations between two entangled vector beams
measured, in the radial and azimuthal DOF respectively, from 16 different polarization measurements.  
Two partially tuned $q$-plates ($\delta_s=\delta_i=\pi/2$) with topological charges $q_i = 1/2$ and $q_s = 1/2$ are used to generate the entangled vector vortex photons.
(c) Continuous exposure images recorded on the camera.
(d) The concurrence of the entangled state and the average concurrence is measured to be $0.556\pm0.079$.
(e) Reconstructed probabilities for the four Bell states $\{\ket{\psi^+},\ket{\psi^-},\ket{\phi^+},\ket{\phi^-}\}$. The insets are the respective theoretical probabilities.
}
\label{correlation1/2&1/2det}
\end{figure*}

\begin{figure*}[tbph!]
\includegraphics [width=0.9\linewidth]{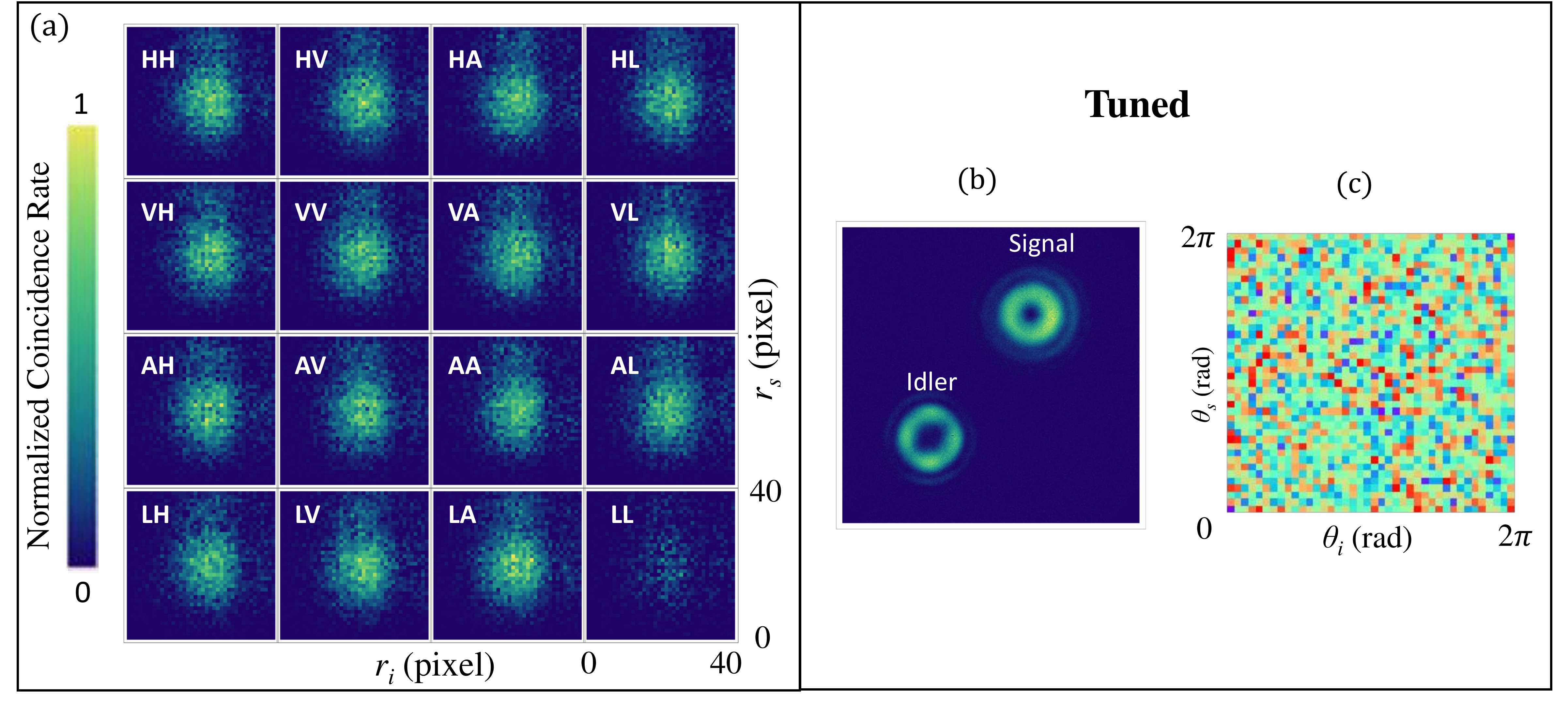}
\centering
\caption{\textbf{Measured spatial correlations and entanglement characterization for bi-photon with spatially varying polarization states generated via two $q$-plates.} 
(a) Two-photon spatial correlations between two entangled vector beams
measured, in the radial coordinate respectively, from 16 different polarization measurements.  
Two tuned $q$-plates ($\delta_s=\delta_i=\pi$) with topological charges $q_i = 1$ and $q_s = 1/2$ are used to generate the entangled vector vortex photons.
(b) Continuous exposure images recorded on the camera.
(c) The concurrence of the entangled state and the average concurrence is measured to be $0.540\pm0.005$.
}
\label{correlation1&1/2}
\end{figure*}

\begin{figure*}[tbph!]
\includegraphics [width=0.9\linewidth]{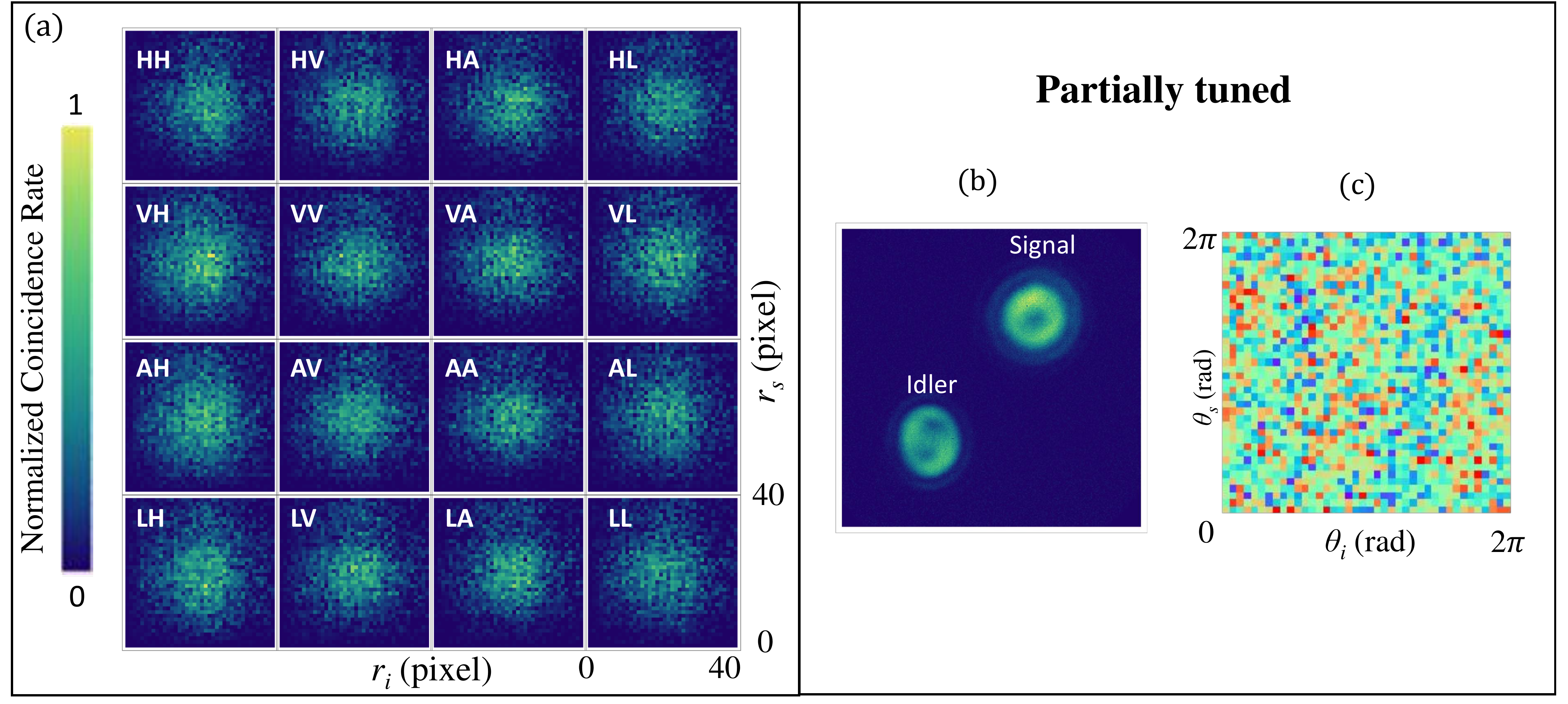}
\centering
\caption{\textbf{Measured spatial correlations and entanglement characterization for bi-photon with spatially varying polarization states generated via two $q$-plates.} 
(a) Two-photon spatial correlations between two entangled vector beams
measured, in the radial coordinate respectively, from 16 different polarization measurements.  
Two partially tuned $q$-plates ($\delta_s=\delta_i=\pi/2$) with topological charges $q_i = 1$ and $q_s = 1/2$ are used to generate the entangled vector vortex photons.
(b) Continuous exposure images recorded on the camera.
(c) The concurrence of the entangled state and the average concurrence is measured to be $0.517\pm0.005$.
}
\label{correlation1&1/2det}
\end{figure*}

\begin{figure*}[tbph!]
\includegraphics [width=0.9\linewidth]{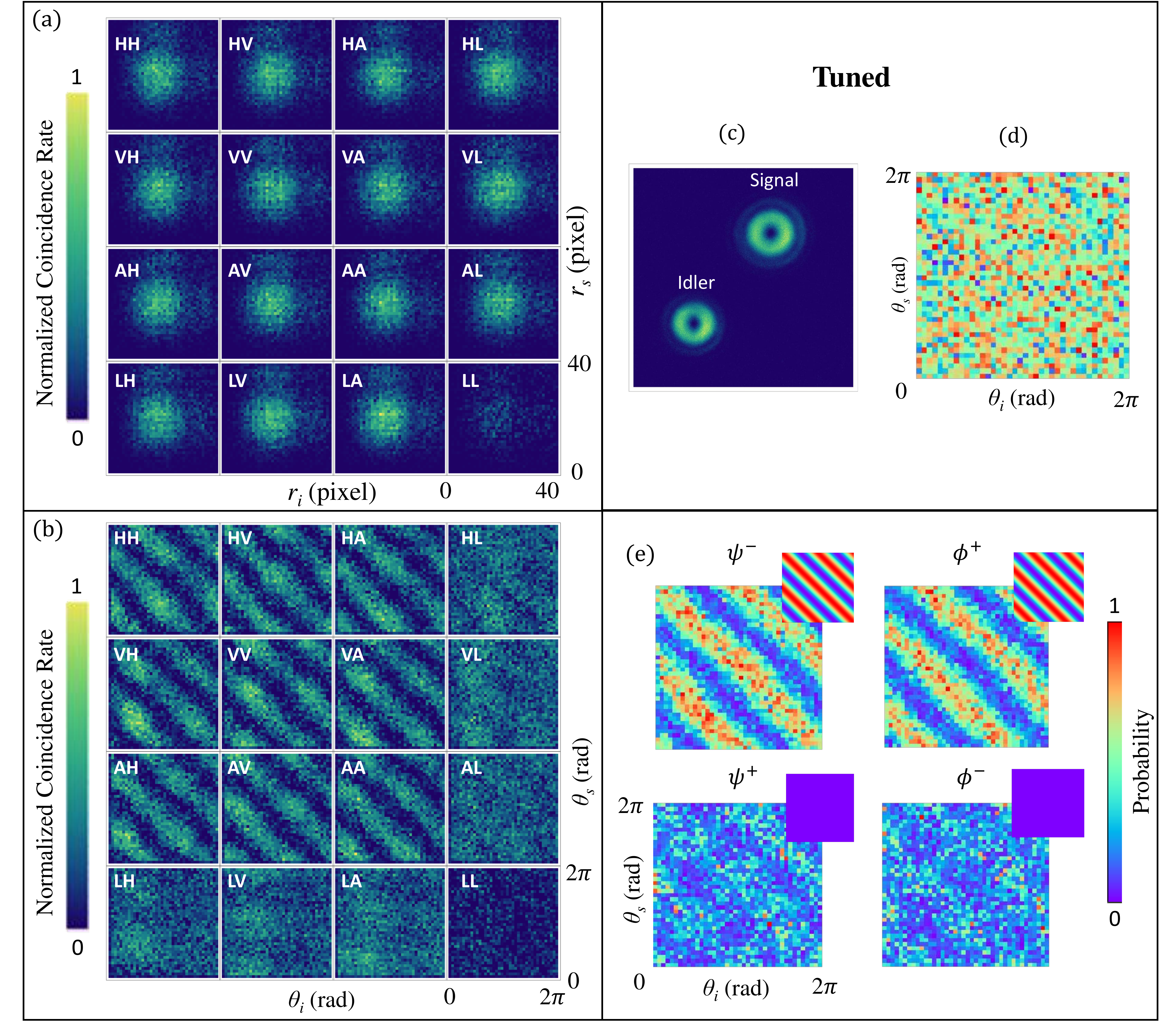}
\centering
\caption{\textbf{Measured spatial correlations and entanglement characterization for bi-photon with spatially varying polarization states generated via two $q$-plates.} 
(a)-(b) Two-photon spatial correlations between two entangled vector beams
measured, in the radial and azimuthal DOF respectively, from 16 different polarization measurements.  
Two tuned $q$-plates ($\delta_s=\delta_i=\pi$) with topological charges $q_i = -1/2$ and $q_s = 1/2$ are used to generate the entangled vector vortex photons.
(c) Continuous exposure images recorded on the camera.
(d) The concurrence of the entangled state and the average concurrence is measured to be $0.575\pm0.010$.
(e) Reconstructed probabilities for the four Bell states $\{\ket{\psi^+},\ket{\psi^-},\ket{\phi^+},\ket{\phi^-}\}$. The insets are the respective theoretical probabilities.
}
\label{correlation-1/2&1/2}
\end{figure*}

\begin{figure*}[tbph!]
\includegraphics [width=0.9\linewidth]{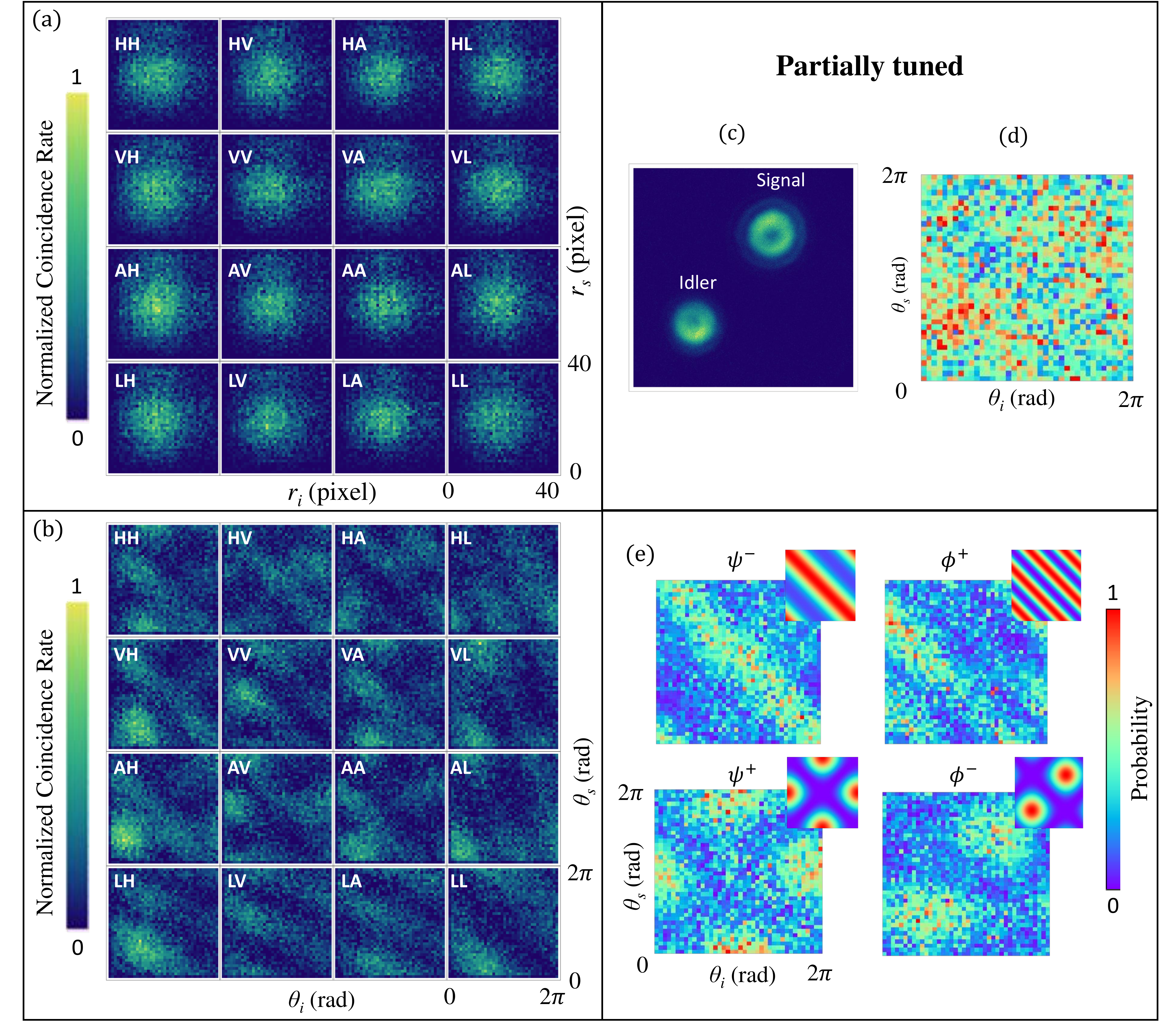}
\centering
\caption{\textbf{Measured spatial correlations and entanglement characterization for bi-photon with spatially varying polarization states generated via two $q$-plates.} 
(a)-(b) Two-photon spatial correlations between two entangled vector beams
measured, in the radial and azimuthal DOF respectively, from 16 different polarization measurements.  
Two partially tuned $q$-plates ($\delta_s=\delta_i=\pi/2$) with topological charges $q_i = -1/2$ and $q_s = 1/2$ are used to generate the entangled vector vortex photons.
(c) Continuous exposure images recorded on the camera.
(d) The concurrence of the entangled state and the average concurrence is measured to be $0.519\pm0.006$.
(e) Reconstructed probabilities for the four Bell states $\{\ket{\psi^+},\ket{\psi^-},\ket{\phi^+},\ket{\phi^-}\}$. The insets are the respective theoretical probabilities.
}
\label{correlation-1/2&1/2det}
\end{figure*}

\begin{figure*}[tbph!]
\includegraphics [width=0.9\linewidth]{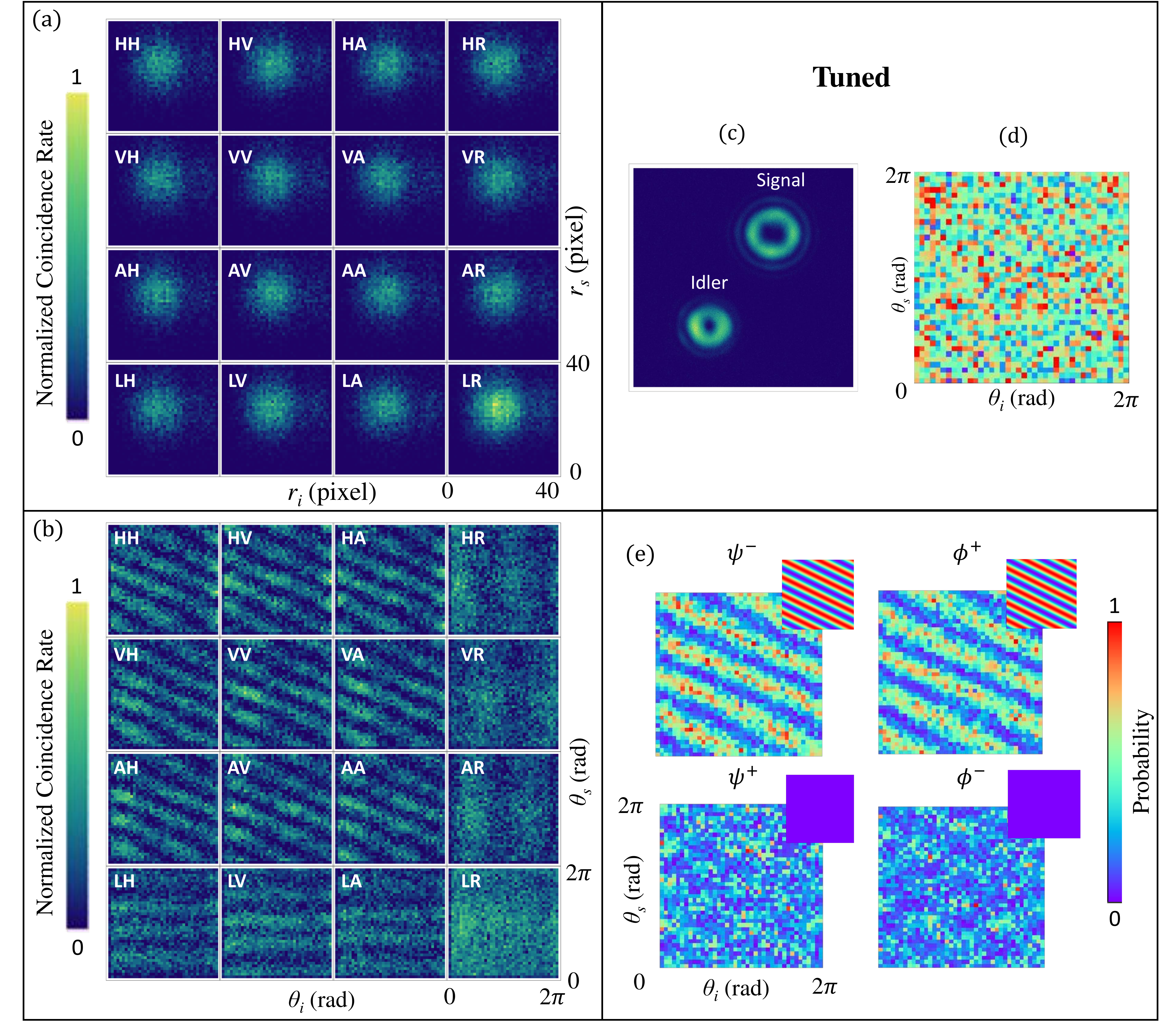}
\centering
\caption{\textbf{Measured spatial correlations and entanglement characterization for bi-photon with spatially varying polarization states generated via two $q$-plates.} 
(a)-(b) Two-photon spatial correlations between two entangled vector beams
measured, in the radial and azimuthal DOF respectively, from 16 different polarization measurements.  
Two tuned $q$-plates ($\delta_s=\delta_i=\pi$) with topological charges $q_i = -1/2$ and $q_s = 1$ are used to generate the entangled vector vortex photons.
(c) Continuous exposure images recorded on the camera.
(d) The concurrence of the entangled state and the average concurrence is measured to be $0.548\pm0.018$.
(e) Reconstructed probabilities for the four Bell states $\{\ket{\psi^+},\ket{\psi^-},\ket{\phi^+},\ket{\phi^-}\}$. The insets are the respective theoretical probabilities.
}
\label{correlation-1/2&1}
\end{figure*}

\begin{figure*}[tbph!]
\includegraphics [width=0.9\linewidth]{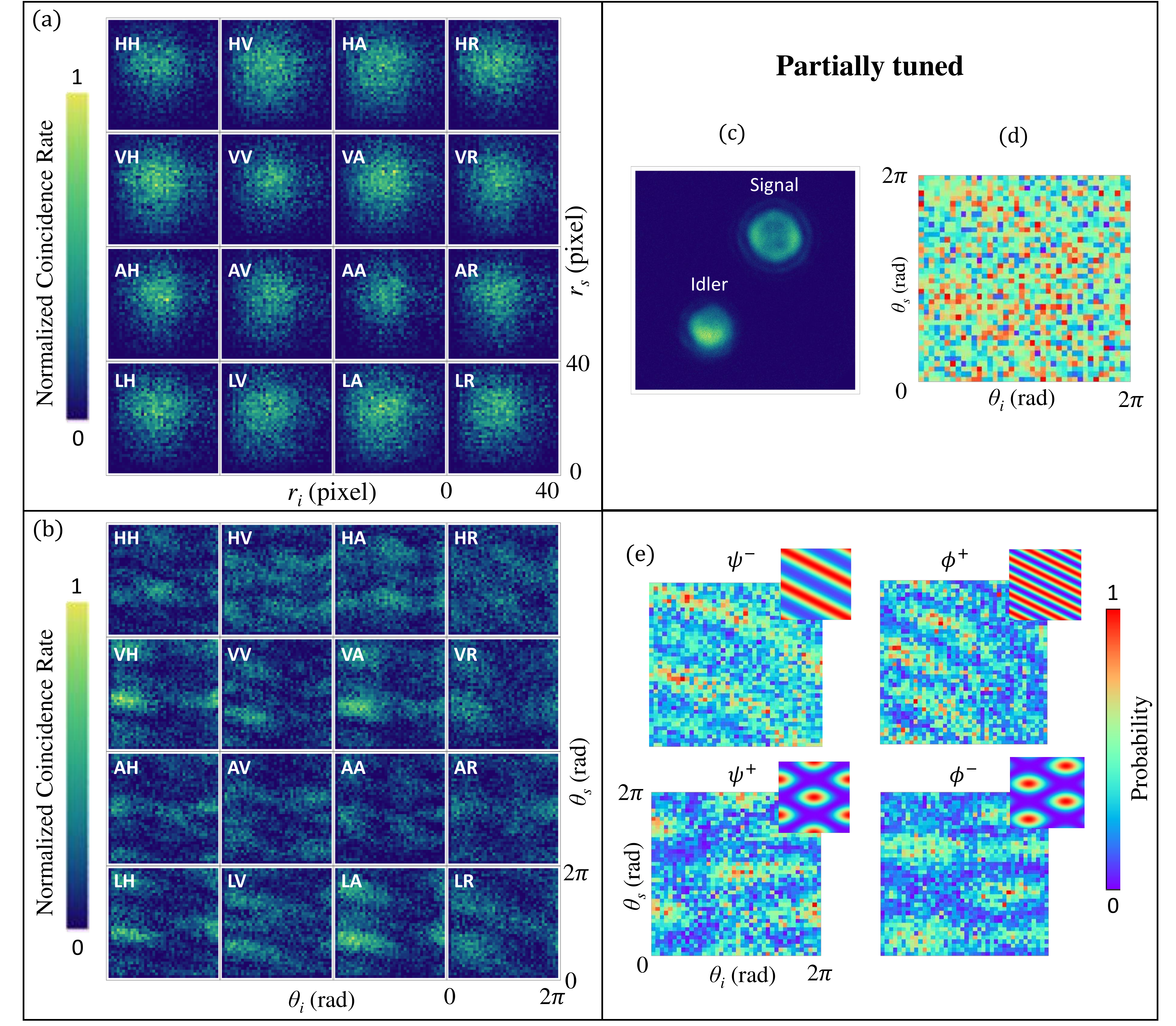}
\centering
\caption{\textbf{Measured spatial correlations and entanglement characterization for bi-photon with spatially varying polarization states generated via two $q$-plates.} 
(a)-(b) Two-photon spatial correlations between two entangled vector beams
measured, in the radial and azimuthal DOF respectively, from 16 different polarization measurements.  
Two partially tuned $q$-plates ($\delta_s=\delta_i=\pi/2$) with topological charges $q_i = -1/2$ and $q_s = 1$ are used to generate the entangled vector vortex photons.
(c) Continuous exposure images recorded on the camera.
(d) The concurrence of the entangled state and the average concurrence is measured to be $0.525\pm0.010$.
(e) Reconstructed probabilities for the four Bell states $\{\ket{\psi^+},\ket{\psi^-},\ket{\phi^+},\ket{\phi^-}\}$. The insets are the respective theoretical probabilities.
}
\label{correlation-1/2&1det}
\end{figure*}

\begin{figure*}[tbph!]
\includegraphics [width=0.9\linewidth]{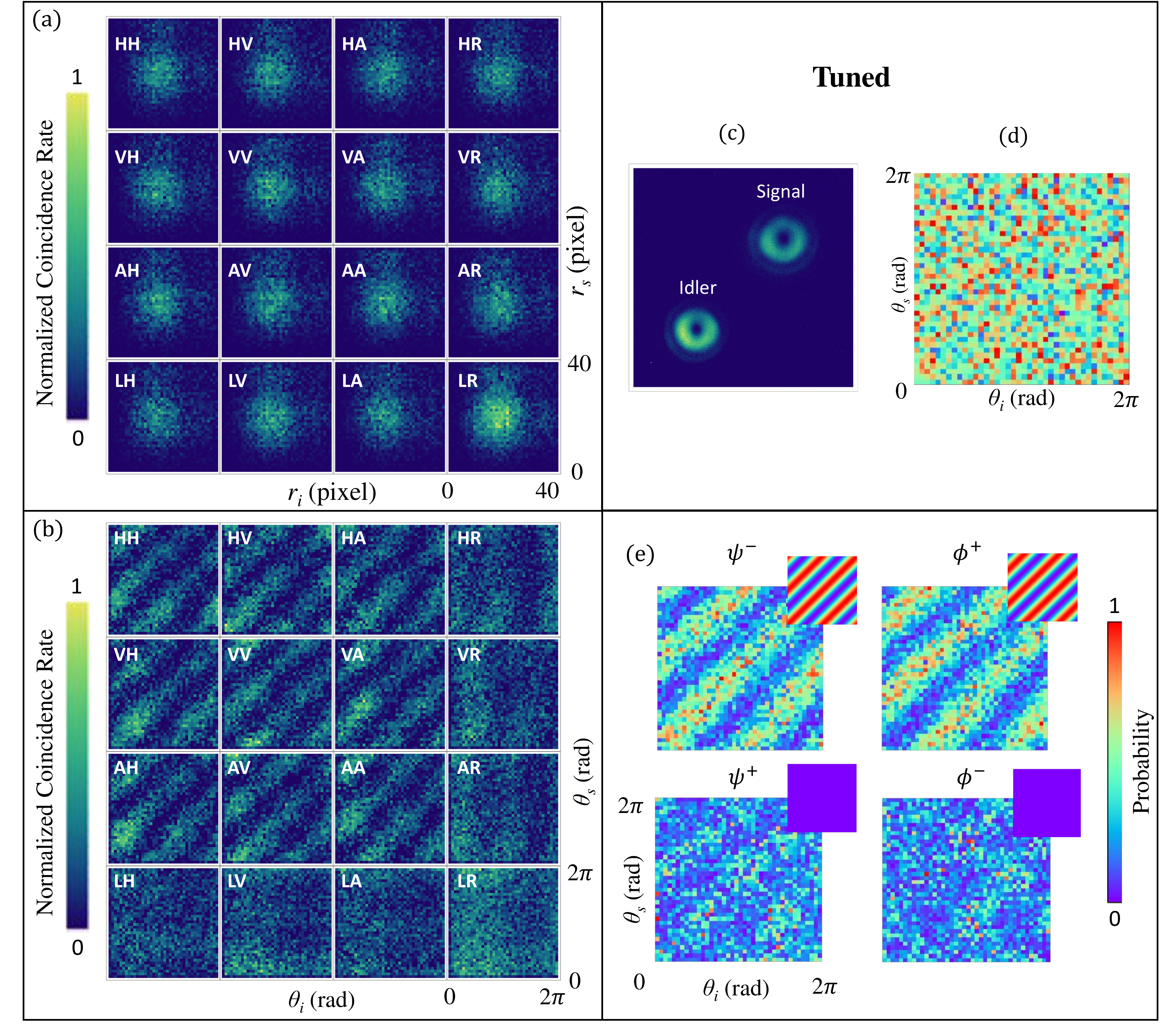}
\centering
\caption{\textbf{Measured spatial correlations and entanglement characterization for bi-photon with spatially varying polarization states generated via two $q$-plates.} 
(a)-(b) Two-photon spatial correlations between two entangled vector beams
measured, in the radial and azimuthal DOF respectively, from 16 different polarization measurements.  
Two tuned $q$-plates ($\delta_s=\delta_i=\pi$) with topological charges $q_i = -1/2$ and $q_s = -1/2$ are used to generate the entangled vector vortex photons.
(c) Continuous exposure images recorded on the camera.
(d) The concurrence of the entangled state and the average concurrence is measured to be $0.555\pm0.005$.
(e) Reconstructed probabilities for the four Bell states $\{\ket{\psi^+},\ket{\psi^-},\ket{\phi^+},\ket{\phi^-}\}$. The insets are the respective theoretical probabilities.
}
\label{correlation-1/2&-1/2}
\end{figure*}

\begin{figure*}[tbph!]
\includegraphics [width=1\linewidth]{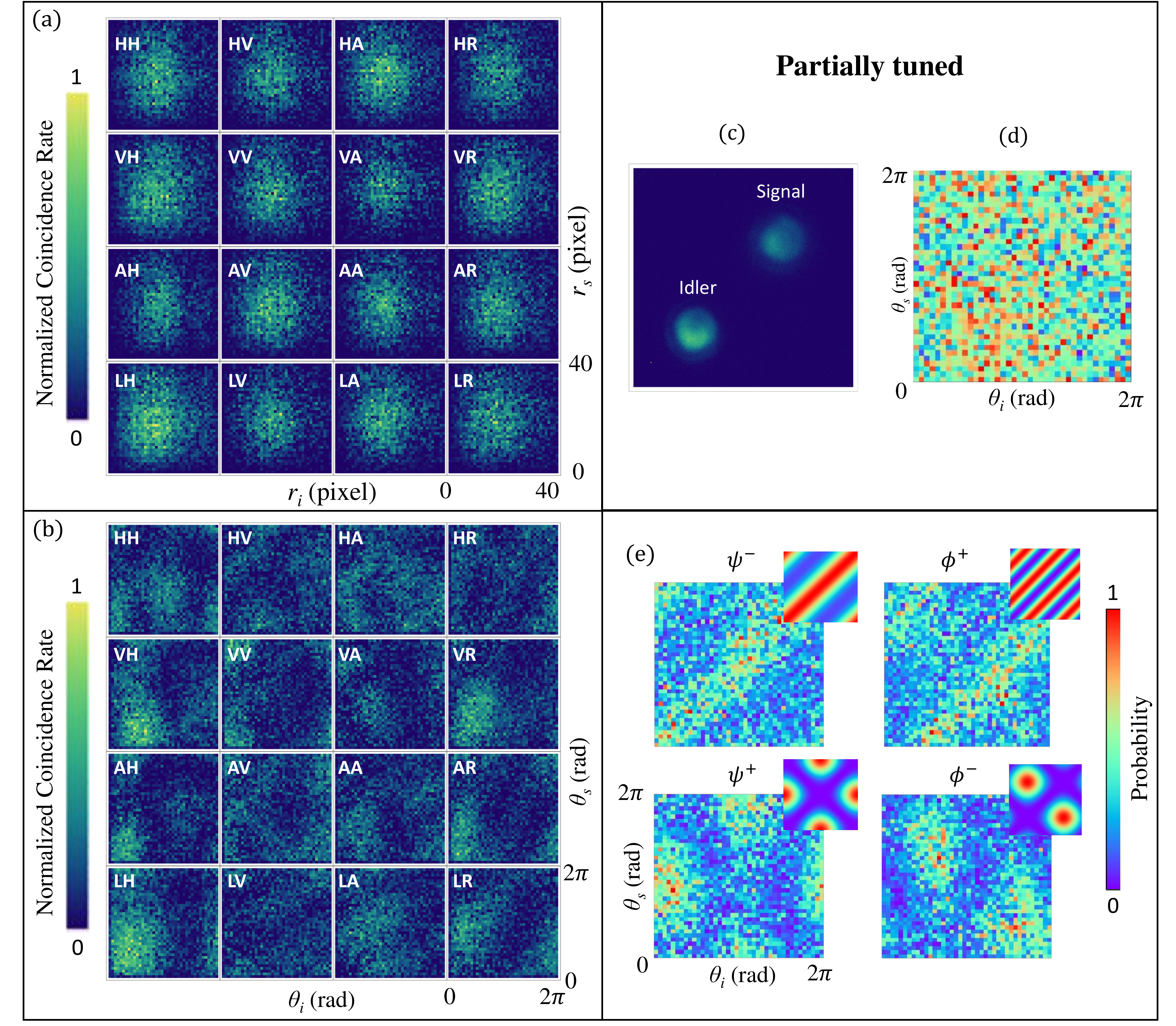}
\centering
\caption{\textbf{Measured spatial correlations and entanglement characterization for bi-photon with spatially varying polarization states generated via two $q$-plates.} 
(a)-(b) Two-photon spatial correlations between two entangled vector beams
measured, in the radial and azimuthal DOF respectively, from 16 different polarization measurements.  
Two partially tuned $q$-plates ($\delta_s=\delta_i=\pi/2$) with topological charges $q_i = -1/2$ and $q_s = -1/2$ are used to generate the entangled vector vortex photons.
(c) Continuous exposure images recorded on the camera.
(d) The concurrence of the entangled state and the average concurrence is measured to be $0.547\pm0.061$.
(e) Reconstructed probabilities for the four Bell states $\{\ket{\psi^+},\ket{\psi^-},\ket{\phi^+},\ket{\phi^-}\}$. The insets are the respective theoretical probabilities.
}
\label{correlation-1/2&-1/2det}
\end{figure*}

\begin{figure*}[tbph!]
\includegraphics [width=0.9\linewidth]{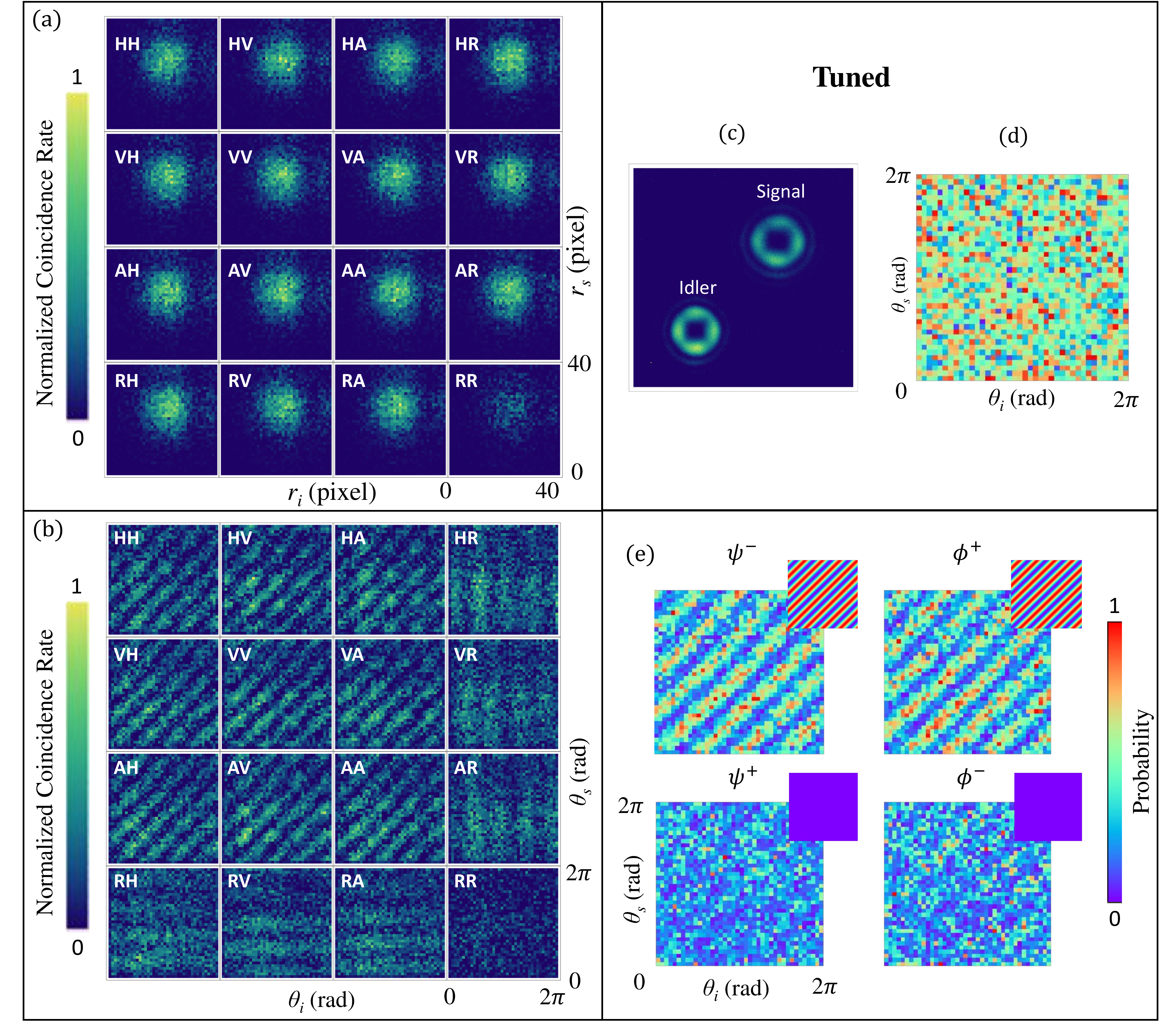}
\centering
\caption{\textbf{Measured spatial correlations and entanglement characterization for bi-photon with spatially varying polarization states generated via two $q$-plates.} 
(a)-(b) Two-photon spatial correlations between two entangled vector beams
measured, in the radial and azimuthal DOF respectively, from 16 different polarization measurements.  
Two tuned $q$-plates ($\delta_s=\delta_i=\pi$) with topological charges $q_i = 1$ and $q_s = 1$ are used to generate the entangled vector vortex photons.
(c) Continuous exposure images recorded on the camera.
(d) The concurrence of the entangled state and the average concurrence is measured to be $0.549\pm0.032$.
(e) Reconstructed probabilities for the four Bell states $\{\ket{\psi^+},\ket{\psi^-},\ket{\phi^+},\ket{\phi^-}\}$. The insets are the respective theoretical probabilities.
}
\label{correlation1&1}
\end{figure*}

\begin{figure*}[tbph!]
\includegraphics [width=1\linewidth]{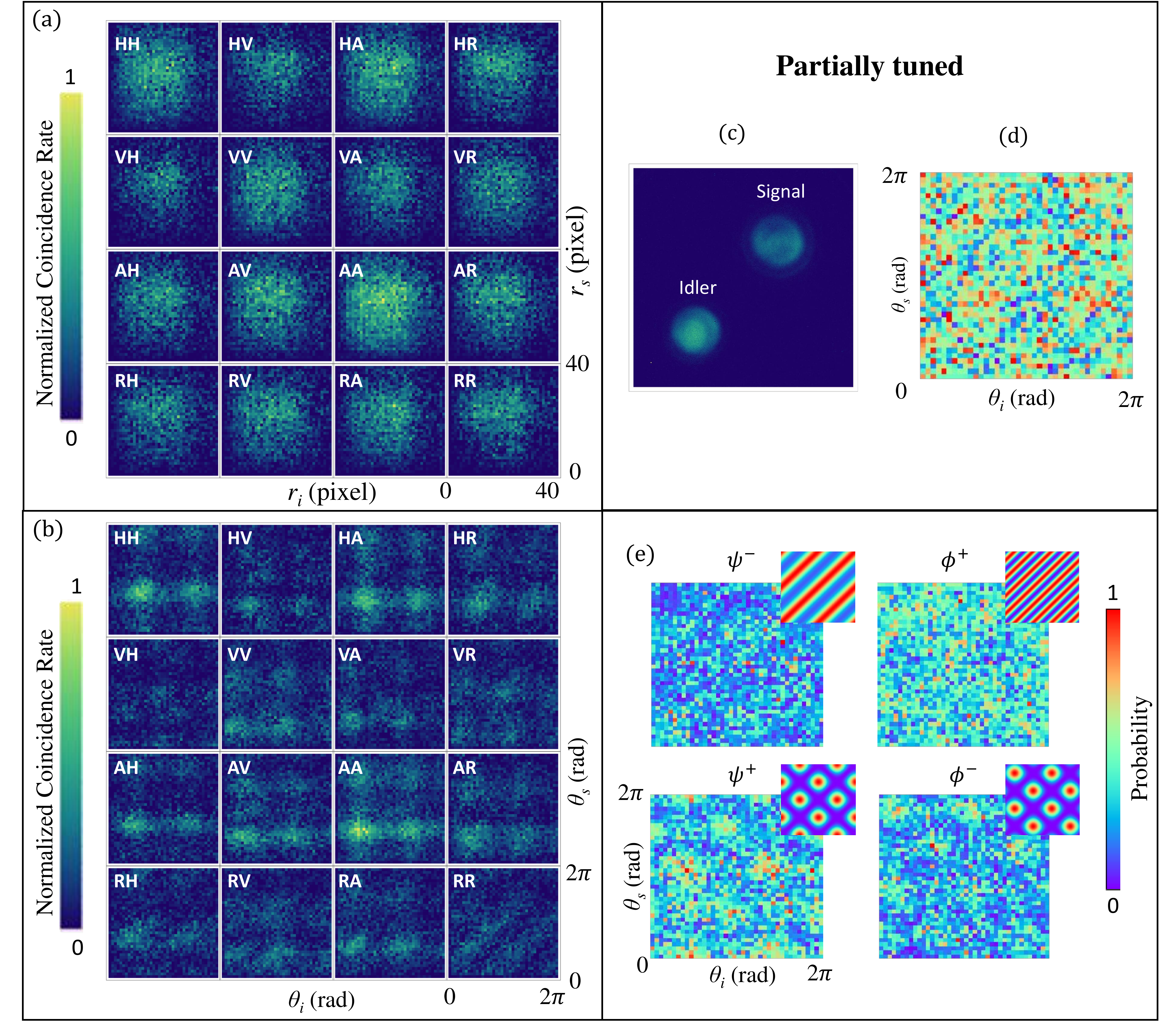}
\centering
\caption{\textbf{Measured spatial correlations and entanglement characterization for bi-photon with spatially varying polarization states generated via two $q$-plates.} 
(a)-(b) Two-photon spatial correlations between two entangled vector beams
measured, in the radial and azimuthal DOF respectively, from 16 different polarization measurements.  
Two partially tuned $q$-plates ($\delta_s=\delta_i=\pi/2$) with topological charges $q_i = 1$ and $q_s = 1$ are used to generate the entangled vector vortex photons.
(c) Continuous exposure images recorded on the camera.
(d) The concurrence of the entangled state and the average concurrence is measured to be $0.518\pm0.013$.
(e) Reconstructed probabilities for the four Bell states $\{\ket{\psi^+},\ket{\psi^-},\ket{\phi^+},\ket{\phi^-}\}$. The insets are the respective theoretical probabilities.
}
\label{correlation1&1det}
\end{figure*}

To verify and characterize the entanglement with vector beams, we measured the correlations in the radial and angular coordinates between two structured beams and observed the generated state changes angularly among the four Bell states, as shown in Figs. \ref{correlation1/2&1/2} - \ref{correlation1&1det}.


\end{document}


\title{Spatially dependant state characterization of entangled vector beams}

\author{Xiaoqin Gao}
\affiliation{Nexus for Quantum Technologies, University of Ottawa, K1N 5N6, Ottawa, ON, Canada}

\author{Yingwen Zhang}
\email{yzhang6@uottawa.ca}
\affiliation{Nexus for Quantum Technologies, University of Ottawa, K1N 5N6, Ottawa, ON, Canada}
\affiliation{National Research Council of Canada, 100 Sussex Drive, K1A 0R6, Ottawa, ON, Canada}

\author{Alessio  D'Errico}
\affiliation{Nexus for Quantum Technologies, University of Ottawa, K1N 5N6, Ottawa, ON, Canada}

\author{Alicia  Sit}
\affiliation{Nexus for Quantum Technologies, University of Ottawa, K1N 5N6, Ottawa, ON, Canada}

\author{Khabat Heshami}
\affiliation{National Research Council of Canada, 100 Sussex Drive, K1A 0R6, Ottawa, ON, Canada}
\affiliation{Nexus for Quantum Technologies, University of Ottawa, K1N 5N6, Ottawa, ON, Canada}

\author{Ebrahim Karimi}
\affiliation{Nexus for Quantum Technologies, University of Ottawa, K1N 5N6, Ottawa, ON, Canada}
\affiliation{National Research Council of Canada, 100 Sussex Drive, K1A 0R6, Ottawa, ON, Canada}


\date{\today}
\maketitle
\clearpage
\onecolumngrid

\section{Two tuned Q-plates}
The unitary action $\hat{U}_q^\delta$ of a $q$-plate with topological charge $q$ acting on a circularly polarized Gaussian beam to first-order approximation can be written as
\begin{equation} \label{eqa1}
    \hat{U}_q\cdot\left[ \begin{array}{ll}
        \ket{L}\\
        \ket{R}\\
    \end{array} \right] = F_{0}(r)\cos\left(\frac{\delta}{2}\right)\left[ \begin{array}{ll}
        \ket{L}\\
        \ket{R}\\
    \end{array} \right]
   + \imath\sin\left(\frac{\delta}{2}\right)F_{q}(r)\left[ \begin{array}{ll}
       e^{-i2q\theta}\ket{R}\\
      e^{i2q\theta}\ket{L}\\
    \end{array} \right],
\end{equation}
where $\ket{L}$ and $\ket{R}$ are left- and right-circular polarizations respectively. 
$F_{q}(r)e^{-i2q\theta}\equiv\text{LG}_{0,2q}(r,\theta)$ being the Laguerre-Gauss mode with orbital angular momentum index $2q$ and radial index zero at $z=0$. $\delta \in [0,\pi]$ is an adjustable parameter that can be tuned based on the voltage applied to the Q-plate. The Q-plates are considered tuned with $\delta=\pi$, and when $\delta=\pi/2$ the Q-plate is considered perfectly detuned.

In our work, we first prepare an anti-symmetric entangled state in the polarization degree of freedom,
\begin{equation} \label{eqa2}
\begin{split}
\ket{\Psi_{0}} & =\frac{1}{\sqrt{2}}(\ket{H}_s\ket{V}_i-\ket{V}_s\ket{H}_i)\\
    & = \frac{\imath}{\sqrt{2}}(\ket{L}_s\ket{R}_i-\ket{R}_s\ket{L}_i),
\end{split}
\end{equation}
with $s$ and $i$ denoting the signal and idler photon respectively. 

Applying two tuned Q-plates ($\delta=\pi$) with topological charges $q_s$ and $q_i$ respectively to each photon gives the following final state,
\begin{equation} \label{eqa3}
\ket{\Psi_{f}} =\frac{\imath}{\sqrt{2}}F_{q_s,q_i}(r_s,r_i)\left[ e^{-2\imath (q_s \theta_s-q_i \theta_i)}\ket{R}_s \ket{L}_i-  e^{2\imath (q_s \theta_s-q_i \theta_i)}\ket{L}_s \ket{R}_i\right],
\end{equation}
where $F_{q_s,q_i}(r_s,r_i)=F_{q_s}(r_s)F_{q_i}(r_i)$. 

Rewriting in the horizontal (H) and vertical (V) polarization basis gives
\begin{equation} \label{eqa3}
\ket{\Psi_{f}} =F_{q_s,q_i}(r_s,r_i)\left[\sin 2(q_s \theta_s-q_i \theta_i)\ket{\phi^+} - \cos 2(q_s \theta_s-q_i \theta_i)\ket{\psi^-}\right].
\end{equation}
where $\ket{\phi^+} = \frac{1}{\sqrt{2}}(\ket{H}_s \ket{H}_i + \ket{V}_s \ket{V}_i$) and $\ket{\psi^-} = \frac{1}{\sqrt{2}}(\ket{H}_s \ket{V}_i - \ket{V}_s \ket{H}_i)$ are two of the four Bell states.

Therefore, the probability for the final state to be in one of the four Bell states are as follows,
\begin{align}
    P_{\phi^+} &= F^2_{q_s,q_i}(r_s,r_i)\sin^2 2(q_s \theta_s-q_i \theta_i) \nonumber\\
    P_{\phi^-} &= 0 \nonumber\\
    P_{\psi^+} &= 0 \nonumber\\
    P_{\psi^-} &= F^2_{q_s,q_i}(r_s,r_i)\cos^2 2(q_s \theta_s-q_i \theta_i)
\end{align}


\section{Two perfectly detuned Q-plates}

When using two perfectly detuned Q-plates ($\delta=\pi/2$), the initial entangled state from eq.\ref{eqa2} becomes,
\begin{align} \label{eqb2}
\ket{\Psi_{f}} & = \frac{i}{2\sqrt{2}} \Bigl\{(F_{0,0}(r_s,r_i)-F_{q_s,q_i}(r_s,r_i)e^{2i(q_s \theta_s-q_i \theta_i})\ket{L}_s\ket{R}_i\nonumber\\
&\quad - \bigl[F_{0,0}(r_s,r_i)-F_{q_s,q_i}(r_s,r_i)e^{-2i(q_s \theta_s-q_i \theta_i}\bigr]\ket{R}_s\ket{L}_i\nonumber\\
&\quad + \bigl[F_{0,q_i}(r_s,r_i)(r_i)e^{2iq_i\theta_i} - F_{q_s,0}(r_s,r_i) e^{2iq_s\theta_s}\bigr]\ket{L}_s\ket{L}_i\nonumber\\
&\quad - \bigl[F_{0,q_i}(r_s,r_i)(r_i)e^{-2iq_i\theta_i} - F_{q_s,0}(r_s,r_i) e^{-2iq_s\theta_s}\bigr]\ket{R}_s\ket{R}_i\Bigr\}.
\end{align}

Rewriting in the H and V basis gives
\begin{align} \label{eqb2}
\ket{\Psi_{f}} & = \frac{1}{2}\Bigl\{-F_{q_s,q_i}(r_s,r_i)\sin{2(q_s \theta_s-q_i \theta_i})\ket{\phi^+}\nonumber\\
&\quad + \bigl[F_{0,0}(r_s,r_i)-F_{q_s,q_i}(r_s,r_i)\sin{2(q_s \theta_s-q_i \theta_i}\bigr]\ket{\psi^-}\nonumber\\
&\quad + \bigl[F_{0,q_i}(r_s,r_i)\sin{2q_i\theta_i} - F_{q_s,0}(r_s,r_i) \sin{2q_s\theta_s}\bigr]\ket{\phi^-}\nonumber\\
&\quad + \bigl[F_{0,q_i}(r_s,r_i)\cos{2q_i\theta_i} - F_{q_s,0}(r_s,r_i) \cos{2q_s\theta_s}\bigr]\ket{\psi^+}\Bigr\}.
\end{align}
with $\ket{\phi^\pm} = \frac{1}{\sqrt{2}}(\ket{H}_s \ket{H}_i \pm \ket{V}_s \ket{V}_i$), and $\ket{\psi^\pm} = \frac{1}{\sqrt{2}}(\ket{H}_s \ket{V}_i \pm \ket{V}_s \ket{H}_i)$ being the four Bell states.

Thus, the probability for the final state to be in one of the four Bell states are as follows,
\begin{align}
    P_{\phi^+} &= \frac{1}{4}F^2_{q_s,q_i}(r_s,r_i)\sin^2 2(q_s \theta_s-q_i \theta_i) \nonumber\\
    P_{\phi^-} &= \frac{1}{4}[F_{0,q_i}(r_s,r_i)\sin{2q_i\theta_i} - F_{q_s,0}(r_s,r_i) \sin{2q_s\theta_s}]^2 \nonumber\\
    P_{\psi^+} &= \frac{1}{4}[F_{0,q_i}(r_s,r_i)\cos{2q_i\theta_i} - F_{q_s,0}(r_s,r_i) \cos{2q_s\theta_s}]^2 \nonumber\\
    P_{\psi^-} &= \frac{1}{4}\left[F_{0,0}(r_s,r_i)-F_{q_s,q_i}(r_s,r_i)\sin{2(q_s \theta_s-q_i \theta_i})\right]^2.
\end{align}

\section{Experimental results}

To verify and characterize the entanglement with vector beams, we measured the correlations in the radial and angular coordinates between two structured beams and observed the generated state changes angularly among the four Bell states, as shown in Figs. \ref{correlation1/2&1/2} - \ref{correlation1&1det}.

%
\begin{figure*}[tbph]
\includegraphics [width= 0.9\textwidth]{half and half.pdf}
\centering
\caption{\textbf{Measured spatial correlations and characterizations for two tuned Q-plates.} 
(a)-(b) The two-photon correlations, in the radial and angular coordinates respectively, were generated by a Q-plate with a fractional topological charge $q_i$ = 1/2 and another Q-plate with a fractional topological charge $q_s$ = 1/2 in each path.
A set of 16 measurements are completed on the 3D camera. ij (i=H, V, A, R, j=H, V, A, L) means different measurement bases. 
(c) Continuous exposure images recorded on the camera.
(d) The concurrence of the entangled state and the average concurrence is measured to be $0.566\pm0.035$.
(e) The experimental results of probabilities of obtaining four Bell states $P_{\phi^{+}}$, $P_{\phi^{-}}$, $P_{\psi^{+}}$, and $P_{\psi^{-}}$.
The insets in the top right are the related theoretical results.
}
\label{correlation1/2&1/2}
\end{figure*}
\begin{figure*}[tbph!]
\includegraphics [width=0.9\linewidth]{half and half det.pdf}
\centering
\caption{\textbf{Measured spatial correlations and characterizations for two perfectly detuned Q-plates.} 
(a)-(b) The two-photon correlations, in the radial and angular coordinates respectively, were generated by a Q-plate with a fractional topological charge $q_i$ = 1/2 and another Q-plate with a fractional topological charge $q_s$ = 1/2 in each path.
A set of 16 measurements are completed on the 3D camera. ij (i=H, V, A, R, j=H, V, A, L) means different measurement bases. 
(c) Continuous exposure images recorded on the camera.
(d) The concurrence of the entangled state and the average concurrence is measured to be $0.556\pm0.079$.
(e) The experimental results of probabilities of obtaining four Bell states $P_{\phi^{+}}$, $P_{\phi^{-}}$, $P_{\psi^{+}}$, and $P_{\psi^{-}}$.
The insets in the top right are the related theoretical results.
}
\label{correlation1/2&1/2det}
\end{figure*}

\begin{figure*}[tbph!]
\includegraphics [width=0.9\linewidth]{half and one.pdf}
\centering
\caption{\textbf{Measured spatial correlations and characterizations for two tuned Q-plates.} 
(a) The two-photon correlations, in the radial coordinates, were generated by a Q-plate with a fractional topological charge $q_i$ = 1 and another Q-plate with a fractional topological charge $q_s$ = 1/2 in each path.
A set of 16 measurements are completed on the 3D camera. ij (i=H, V, A, L, j=H, V, A, L) means different measurement bases.
(b) Continuous exposure images recorded on the camera.
(c) The concurrence of the entangled state and the average concurrence is measured to be $0.540\pm0.005$.
}
\label{correlation1&1/2}
\end{figure*}

\begin{figure*}[tbph!]
\includegraphics [width=0.9\linewidth]{half and one det.pdf}
\centering
\caption{\textbf{Measured spatial correlations and characterizations for two perfectly detuned Q-plates.} 
(a) The two-photon correlations, in the radial coordinates, were generated by a Q-plate with a fractional topological charge $q_i$ = 1 and another Q-plate with a fractional topological charge $q_s$ = 1/2 in each path.
A set of 16 measurements are completed on the 3D camera. ij (i=H, V, A, L, j=H, V, A, L) means different measurement bases.
(b) Continuous exposure images recorded on the camera.
(c) The concurrence of the entangled state and the average concurrence is measured to be $0.517\pm0.005$.
}
\label{correlation1&1/2det}
\end{figure*}

\begin{figure*}[tbph!]
\includegraphics [width=0.9\linewidth]{-half and half.pdf}
\centering
\caption{\textbf{Measured spatial correlations and characterizations for two tuned Q-plates.} 
(a)-(b) The two-photon correlations, in the radial and angular coordinates respectively, were generated by a Q-plate with a fractional topological charge $q_i$ = -1/2 and another Q-plate with a fractional topological charge $q_s$ = 1/2 in each path.
A set of 16 measurements are completed on the 3D camera. ij (i=H, V, A, L, j=H, V, A, L) means different measurement bases.
(c) Continuous exposure images recorded on the camera.
(d) The concurrence of the entangled state and the average concurrence is measured to be $0.575\pm0.010$.
(e) The experimental results of probabilities of obtaining four Bell states $P_{\phi^{+}}$, $P_{\phi^{-}}$, $P_{\psi^{+}}$, and $P_{\psi^{-}}$.
The insets in the top right are the related theoretical results.
}
\label{correlation-1/2&1/2}
\end{figure*}

\begin{figure*}[tbph!]
\includegraphics [width=0.9\linewidth]{-half and half det.pdf}
\centering
\caption{\textbf{Measured spatial correlations and characterizations for two perfectly detuned Q-plates.} 
(a)-(b) The two-photon correlations, in the radial and angular coordinates respectively, were generated by a Q-plate with a fractional topological charge $q_i$ = -1/2 and another Q-plate with a fractional topological charge $q_s$ = 1/2 in each path.
A set of 16 measurements are completed on the 3D camera. ij (i=H, V, A, L, j=H, V, A, L) means different measurement bases.
(c) Continuous exposure images recorded on the camera.
(d) The concurrence of the entangled state and the average concurrence is measured to be $0.519\pm0.006$.
(e) The experimental results of probabilities of obtaining four Bell states $P_{\phi^{+}}$, $P_{\phi^{-}}$, $P_{\psi^{+}}$, and $P_{\psi^{-}}$.
The insets in the top right are the related theoretical results.
}
\label{correlation-1/2&1/2det}
\end{figure*}

\begin{figure*}[tbph!]
\includegraphics [width=0.9\linewidth]{-half and one.pdf}
\centering
\caption{\textbf{Measured spatial correlations and characterizations for two tuned Q-plates.} 
(a)-(b) The two-photon correlations,  in the radial and angular coordinates respectively, were generated by a Q-plate with a fractional topological charge $q_i$ = -1/2 and another Q-plate with a fractional topological charge $q_s$ = 1 in each path.
A set of 16 measurements are completed on the 3D camera. ij (i=H, V, A, L, j=H, V, A, R) means different measurement bases.
(c) Continuous exposure images recorded on the camera.
(d) The concurrence of the entangled state and the average concurrence is measured to be $0.548\pm0.018$.
(e) The experimental results of probabilities of obtaining four Bell states $P_{\phi^{+}}$, $P_{\phi^{-}}$, $P_{\psi^{+}}$, and $P_{\psi^{-}}$.
The insets in the top right are the related theoretical results.
}
\label{correlation-1/2&1}
\end{figure*}

\begin{figure*}[tbph!]
\includegraphics [width=0.9\linewidth]{-half and one det.pdf}
\centering
\caption{\textbf{Measured spatial correlations and characterizations for two perfectly detuned Q-plates.} 
(a)-(b) The two-photon correlations, in the radial and angular coordinates respectively, were generated by a Q-plate with a fractional topological charge $q_i$ = -1/2 and another Q-plate with a fractional topological charge $q_s$ = 1 in each path.
A set of 16 measurements are completed on the 3D camera. ij (i=H, V, A, L, j=H, V, A, R) means different measurement bases. 
(c) Continuous exposure images recorded on the camera.
(d) The concurrence of the entangled state and the average concurrence is measured to be $0.525\pm0.010$.
(e) The experimental results of probabilities of obtaining four Bell states $P_{\phi^{+}}$, $P_{\phi^{-}}$, $P_{\psi^{+}}$, and $P_{\psi^{-}}$.
The insets in the top right are the related theoretical results.
}
\label{correlation-1/2&1det}
\end{figure*}

\begin{figure*}[tbph!]
\includegraphics [width=0.9\linewidth]{-half and -half.pdf}
\centering
\caption{\textbf{Measured spatial correlations and characterizations for two tuned Q-plates.}
(a)-(b) The two-photon correlations, in the radial and angular coordinates respectively, were generated by a Q-plate with a fractional topological charge $q_i$ = -1/2 and another Q-plate with a fractional topological charge $q_s$ = -1/2 in each path.
A set of 16 measurements are completed on the 3D camera. ij (i=H, V, A, L, j=H, V, A, R) means different measurement bases.
(c) Continuous exposure images recorded on the camera.
(d) The concurrence of the entangled state and the average concurrence is measured to be $0.555\pm0.005$.
(e) The experimental results of probabilities of obtaining four Bell states $P_{\phi^{+}}$, $P_{\phi^{-}}$, $P_{\psi^{+}}$, and $P_{\psi^{-}}$.
The insets in the top right are the related theoretical results.
}
\label{correlation-1/2&-1/2}
\end{figure*}

\begin{figure*}[tbph!]
\includegraphics [width=1\linewidth]{-half and -half det.pdf}
\centering
\caption{\textbf{Measured spatial correlations and characterizations for two perfectly detuned Q-plates.} 
(a)-(b) The two-photon correlations, in the radial and angular coordinates respectively, were generated by a Q-plate with a fractional topological charge $q_i$ = -1/2 and another Q-plate with a fractional topological charge $q_s$ =-1/2 in each path.
A set of 16 measurements are completed on the 3D camera. ij (i=H, V, A, L, j=H, V, A, R) means different measurement bases. 
(c) Continuous exposure images recorded on the camera.
(d) The concurrence of the entangled state and the average concurrence is measured to be $0.547\pm0.061$.
(e) The experimental results of probabilities of obtaining four Bell states $P_{\phi^{+}}$, $P_{\phi^{-}}$, $P_{\psi^{+}}$, and $P_{\psi^{-}}$.
The insets in the top right are the related theoretical results.
}
\label{correlation-1/2&-1/2det}
\end{figure*}

\begin{figure*}[tbph!]
\includegraphics [width=0.9\linewidth]{one and one.pdf}
\centering
\caption{\textbf{Measured spatial correlations and characterizations for two tuned Q-plates.}
(a)-(b) The two-photon correlations, in the radial and angular coordinates respectively, were generated by a Q-plate with a fractional topological charge $q_i$ = 1 and another Q-plate with a fractional topological charge $q_s$ = 1 in each path.
A set of 16 measurements are completed on the 3D camera. ij (i=H, V, A, R, j=H, V, A, R) means different measurement bases.
(c) Continuous exposure images recorded on the camera.
(d) The concurrence of the entangled state and the average concurrence is measured to be $0.549\pm0.032$.
(e) The experimental results of probabilities of obtaining four Bell states $P_{\phi^{+}}$, $P_{\phi^{-}}$, $P_{\psi^{+}}$, and $P_{\psi^{-}}$.
The insets in the top right are the related theoretical results.
}
\label{correlation1&1}
\end{figure*}

\begin{figure*}[tbph!]
\includegraphics [width=1\linewidth]{one and one det.pdf}
\centering
\caption{\textbf{Measured spatial correlations and characterizations for two perfectly detuned Q-plates.} 
(a)-(b) The two-photon correlations, in the radial and angular coordinates respectively, were generated by a Q-plate with a fractional topological charge $q_i$ = 1 and another Q-plate with a fractional topological charge $q_s$ = 1 in each path.
A set of 16 measurements are completed on the 3D camera. ij (i=H, V, A, R, j=H, V, A, R) means different measurement bases. 
(c) Continuous exposure images recorded on the camera.
(d) The concurrence of the entangled state and the average concurrence is measured to be $0.518\pm0.013$.
(e) The experimental results of probabilities of obtaining four Bell states $P_{\phi^{+}}$, $P_{\phi^{-}}$, $P_{\psi^{+}}$, and $P_{\psi^{-}}$.
The insets in the top right are the related theoretical results.
}
\label{correlation1&1det}
\end{figure*}